\documentclass[aps,prd,twocolumn,showkeys,amsmath,amssymb]{revtex4-1}
\usepackage{graphicx}
\usepackage{epstopdf}
\usepackage{multirow}
\usepackage{soul}
\usepackage{dcolumn}
\usepackage{bm}
\usepackage{mathrsfs}
\usepackage[colorlinks,citecolor=blue,anchorcolor=red,menucolor=red,linkcolor=red,filecolor=red,runcolor=red,urlcolor=blue,frenchlinks=red]{hyperref}
\usepackage{color}
\usepackage{array}
\usepackage{longtable}
\usepackage{subfigure}
\usepackage{float}

\allowdisplaybreaks[3]

\newcommand{\mev}{\textrm{ MeV}}

\newcommand{\barb}{\bar{B}}

\newcommand{\barbs}{\bar{B}_s}
\newcommand{\ds}{D_s}

\newcommand{\barbss}{\bar{B}_s^*}
\newcommand{\dss}{D_s^*}

\begin{document}

\title{Hadronic molecular states with the quark contents $bc\bar{s}\bar{q}$, $b\bar{c}s\bar{q}$, and $b\bar{c}\bar{s}q$}

\author{Wen-Ying Liu$^1$}
\author{Hua-Xing Chen$^1$}
\email{hxchen@seu.edu.cn}
\author{En Wang$^2$}
\affiliation{$^1$School of Physics, Southeast University, Nanjing 210094, China}
\affiliation{$^2$School of Physics, Zhengzhou University, Zhengzhou 450001, China}
\begin{abstract}
We study the hadronic molecular states with the quark content $bc\bar{s}\bar{q}$ by investigating the interactions of the $\bar{B}_s D$, $\bar{B} D_s$, $\bar{B}_s^* D$, $\bar{B}^* D_s$, $\bar{B}_s D^*$, $\bar{B} D_s^*$, $\bar{B}_s^* D^*$, and $\bar{B}^* D_s^*$ systems. By solving the Bethe-Salpeter equation within the extended local hidden gauge formalism, we find altogether six poles qualifying as possible hadronic molecular states: one pole of $J^P=0^+$ below the $\bar{B}_s D$-$\bar{B}D_s$ threshold, one pole of $J^P=1^+$ below the $\bar{B}_s^* D$-$\bar{B}^* D_s$ threshold, one pole of $J^P=1^+$ below the $\bar{B}_s D^*$-$\bar{B}D_s^*$ threshold, and three poles of $J^P=0^+/1^+/2^+$ below the $\bar{B}_s^* D^*$-$\bar{B}^* D_s^*$ threshold. Their binding energies are calculated to be about 10-20~MeV with the cut-off momentum $q_\textrm{max}=600\textrm{ MeV}$. Similarly, we study the hadronic molecular states with $bs\bar{c}\bar{q}$ by investigating the interactions of the  $\bar{B}\bar{D}_s$, $\bar{B}_c\bar{K}$, $\bar{B}^*\bar{D}_s$, $\bar{B}_c^*\bar{K}$, $\bar{B}\bar{D}_s^*$, $\bar{B}_c\bar{K}^*$, $\bar{B}^*\bar{D}_s^*$, $\bar{B}_c^*\bar{K}^*$ systems, and the states with $bq\bar{c}\bar{s}$ by investigating the interactions of the  $\bar{B}_s\bar{D}$, $\bar{B}_cK$, $\bar{B}_s^*\bar{D}$, $\bar{B}_c^*K$, $\bar{B}_s\bar{D}^*$, $\bar{B}_cK^*$, $\bar{B}_s^*\bar{D}^*$, $\bar{B}_c^*K^*$ systems. However, no deeply-bound poles are found in these systems.
\end{abstract}
%
%
\maketitle
\pagenumbering{arabic}

\section{Introduction}
\label{sec:intro}

Recently, the LHCb Collaboration reported their observation of the first doubly charmed tetraquark state $T_{cc}(3875)$ in the $D^0D^0\pi^+$ mass spectrum just below the $D^{*+}D^0$ mass threshold~\cite{LHCb:2021vvq,LHCb:2021auc}. This state has the quark content $cc\bar u\bar d$. Its spin-parity quantum numbers were determined to be $J^P = 1^+$, and the LHCb experiment favors it to be an isoscalar state. Based on the Breit-Wigner parametrisation, its mass and width were measured to be:
\begin{eqnarray}
M_{\rm BW} &=& M_{D^{*+}} + M_{D^0} - \left( 273 \pm 61 \pm 5 ^{+11}_{-14} \right) {\rm~keV} \, ,
\\ \nonumber \Gamma_{\rm BW} &=& 410 \pm 165 \pm 43 ^{+18}_{-38} {\rm~keV} \, .
\end{eqnarray}
A LHCb analysis of the data with a unitary amplitude and considering the experimental resolution produces the resonance pole at $\sqrt{s} = m_{\rm pole} - {i\over2} \Gamma_{\rm pole}$, where~\cite{LHCb:2021auc}
\begin{eqnarray}
m_{\rm pole} &=& M_{D^{*+}} + M_{D^0} - \left( 360 \pm 40^{+0}_{-4} \right) {\rm~keV} \, ,
\\ \nonumber \Gamma_{\rm pole} &=&  48 \pm 2 ^{+~0}_{ - 14}  {\rm~keV} \, .
\end{eqnarray}
The closeness of the $T_{cc}(3875)$ to the $D^{*+}D^0$ threshold makes it a good candidate for the $DD^*$ hadronic molecular state of $I(J^P) = 0(1^+)$, whose existence had been predicted in Refs.~\cite{Janc:2004qn,Yang:2009zzp,Li:2012ss,Ohkoda:2012hv,Xu:2017tsr,Liu:2019stu,Ding:2020dio,Qin:2020zlg} before the LHCb experiment.

Besides, the BESIII Collaboration observed an excess of events near the $D_s^- D^{*0}$-$D^{*-}_s D^0$ mass thresholds in the $K^+$ recoil-mass spectrum of the $e^+e^- \to K^+(D_s^-D^{*0} + D_s^{*-}D^0)$ process~\cite{BESIII:2020qkh}. This structure, denoted as $Z_{cs}(3985)$, is expected to be the strange partner of the $Z_c(3900)$~\cite{BESIII:2013ris,Belle:2013yex}. Its pole mass and width were measured to be $3982.5^{+1.8}_{-2.6} \pm 2.1 \mev$ and $12.8^{+5.3}_{-4.4} \pm 3.0 \mev$, respectively. It is the first candidate for the hidden-charm tetraquark state with strangeness.

Later the LHCb Collaboration reported their observation of two exotic structures in the $J/\psi K^+$ mass distribution of the $B^+ \to J/\psi\phi K^+$ decay~\cite{LHCb:2021uow}. The mass and width of the lower-lying state, denoted as $Z_{cs}(4000)$, were measured to be $4003 \pm 6 ^{+4}_{-14} \mev$ and $131 \pm 15 \pm 26 \mev$, respectively. Its spin-parity quantum numbers were determined to be $J^P=1^+$. The mass and width of the higher-lying state, denoted as $Z_{cs}(4220)$, were measured to be $4216 \pm 24 ^{+43}_{-30} \mev$ and $233 \pm 52 ^{+97}_{-73} \mev$, respectively. Its spin-parity quantum numbers were determined to be either $J^P=1^+$ or $1^-$.

The above $Z_{cs}$ states have the quark content $c\bar{c}s\bar{q}$ or $c\bar{c}\bar{s}q$ ($q=u/d$). There have been extensive theoretical studies, and their existence had been predicted in various theoretical models before the BESIII and LHCb experiments, based on the $D \bar D_s^*$-$D^* \bar D_s$ hadronic molecular picture~\cite{Lee:2008uy}, the compact tetraquark picture~\cite{Ebert:2005nc,Dias:2013qga}, the hadro-quarkonium picture~\cite{Voloshin:2019ilw,Ferretti:2020ewe}, and the initial-single-chiral-particle-emission mechanism~\cite{Chen:2013wca}. We refer to Refs.~\cite{Chen:2022asf,Chen:2016qju,Liu:2019zoy,Lebed:2016hpi,Esposito:2016noz,Hosaka:2016pey,Guo:2017jvc,Ali:2017jda,Olsen:2017bmm,Karliner:2017qhf,Bass:2018xmz,Brambilla:2019esw,Guo:2019twa,Ketzer:2019wmd,Yang:2020atz,Roberts:2021nhw,Fang:2021wes,Jin:2021vct,JPAC:2021rxu,Meng:2022ozq,Mai:2022eur,Maiani:2022psl,Ling:2021bir,Chen:2020aos,Chen:2020uif,Chen:2021erj} for their detailed discussions, and the studies on the $T_{cc}(3875)$ can also be found in these reviews.

The observation of the $T_{cc}(3875)$ with the quark content $cc\bar u\bar d$ motivates us to investigate the hadronic molecular states with the quark content $bc\bar{q}\bar{q}$, and the observations of the three $Z_{cs}$ states with the quark contents $c\bar{c}s\bar{q}/c\bar{c}\bar{s}q$ motivate us to further investigate the hadronic molecular states with the quark contents $bc\bar{s}\bar{q}$, $b\bar{c}s\bar{q}$, and $b\bar{c}\bar{s}q$. Accordingly, in this paper we shall study the possibly-existing hadronic molecular states with these quark contents. We shall use the extended local hidden gauge symmetry approach~\cite{Meissner:1987ge,Bando:1987br,Nagahiro:2008cv}, which has been widely applied in the study of the meson-meson and meson-baryon interactions~\cite{Wu:2010jy,Duan:2021pll,Zhang:2020rqr,Wang:2017mrt,Geng:2008gx,Xiao:2013yca,Molina:2008jw,Uchino:2015uha}. We also refer to Refs.~\cite{Oller:1997ti,Oller:1998hw,Oller:1997ng,Oset:1997it,Jido:2003cb} for more discussions.

For the possible hadronic molecular states with the quark content $bc\bar{s}\bar{q}$, we shall investigate the interactions of the $\barbs D$, $\barb\ds$, $\barbss D$, $\bar{B}^* D_s$, $\barbs D^*$, $\barb \dss$, $\barbss D^*$, and $\barb^* \dss$ systems. We shall find six poles in these systems, which may qualify as hadronic molecular states. Besides, we shall study the hadronic molecular states with $b\bar{c}s\bar{q}$ by investigating the interactions of the $\bar{B}\bar{D}_s$, $\bar{B}_c\bar{K}$, $\bar{B}^*\bar{D}_s$, $\bar{B}_c^*\bar{K}$, $\bar{B}\bar{D}_s^*$, $\bar{B}_c\bar{K}^*$, $\bar{B}^*\bar{D}_s^*$, $\bar{B}_c^*\bar{K}^*$ systems, and the states with $b\bar{c}\bar{s}q$ by investigating the interactions of the  $\bar{B}_s\bar{D}$, $\bar{B}_cK$, $\bar{B}_s^*\bar{D}$, $\bar{B}_c^*K$, $\bar{B}_s\bar{D}^*$, $\bar{B}_cK^*$, $\bar{B}_s^*\bar{D}^*$, $\bar{B}_c^*K^*$ systems. However, we shall find no deeply-bound pole in these systems.

This paper is organized as follows. In Sec.~\ref{sec:formlism} we apply the local hidden gauge formalism to derive the potentials for the interactions between charmed(-strange) mesons and bottom(-strange) mesons. Based on the obtained potentials, we solve the coupled-channel Bethe-Salpeter equation in Sec.~\ref{sec:result} to extract the poles, some of which can qualify as hadronic molecular states. A brief summary is given in Sec.~\ref{sec:summary}.

\section{Local Hidden Gauge Formalism}
\label{sec:formlism}

By using the unitary coupled-channel approach within the local hidden gauge formalism, the interactions of the $B^{(*)}D^{(*)}$ and $B^{(*)}\bar{D}^{(*)}$ systems have been systematically studied in Ref.~\cite{Sakai:2017avl}, and the interactions of the $B_{(s)}^{(*)}B_{(s)}^{(*)}$ and $B^{(*)}K^{(*)}$ systems have been systematically studied in Refs.~\cite{Dai:2022ulk,Oset:2022xji}. In this section we shall extend these formalisms to the $\bar B_{(s)}^{(*)}D_{(s)}^{(*)}$ and $\bar B_{(s)}^{(*)}\bar D_{(s)}^{(*)}$ systems:
\begin{itemize}

\item We shall investigate the $\bar{B}_s^0D^+$, $\bar{B}^0D_s^+$, $\bar{B}_s^{*0}D^+$, $\bar{B}^{*0}D_s^+$, $\bar{B}_s^0D^{*+}$, $\bar{B}^0D_s^{*+}$, $\bar{B}_s^{*0}D^{*+}$, and $\bar{B}^{*0}D_s^{*+}$ channels to study the hadronic molecular states with the quark content $bc\bar{s}\bar{d}$.

\item We shall investigate the  $\bar{B}^0D_s^-$, $B_c^-\bar{K}^0$, $\bar{B}^{*0}D_s^-$, $B_c^{*-}\bar{K}^0$, $\bar{B}^0D_s^{*-}$, $B_c^-\bar{K}^{*0}$, $\bar{B}^{*0}D_s^{*-}$, and $B_c^{*-}\bar{K}^{*0}$ channels to study those with $b\bar{c}s\bar{d}$.

\item We shall investigate the  $\bar{B}_s^0D^-$, $B_c^-K^0$, $\bar{B}_s^{*0}D^-$, $B_c^{*-}K^0$, $\bar{B}_s^0D^{*-}$, $B_c^-K^{*0}$, $\bar{B}_s^{*0}D^{*-}$, and $B_c^{*-}K^{*0}$ channels to study those with $b\bar{c}\bar{s}d$.

\end{itemize}
The threshold masses of the above channels are tabulated in Table~\ref{tab:threshold}. Besides, we shall also study the hadronic molecular states with the quark contents $bc\bar{s}\bar{u}$, $b\bar{c}s\bar{u}$, and $b\bar{c}\bar{s}u$. It is only the third component of isospin that changes, and the threshold masses of these channels are also tabulated in Table~\ref{tab:threshold}.

\begin{table*}[htb]
\renewcommand{\arraystretch}{1.4}
\centering
\caption{Threshold masses of the 48 channels considered in the present study, in units of MeV.}
\setlength{\tabcolsep}{2mm}{
\begin{tabular}{c|cccccccc}
\hline\hline
Channels & $\bar{B}_s^0D^+$ & $\bar{B}^0D_s^+$ & $\bar{B}_s^{*0}D^+$ & $\bar{B}^{*0}D_s^+$ & $\bar{B}_s^0D^{*+}$ & $\bar{B}^0D_s^{*+}$ & $\bar{B}_s^{*0}D^{*+}$ & $\bar{B}^{*0}D_s^{*+}$
\\ \hline
Threshold & 7236.6 & 7248.0 & 7285.1 & 7293.1 & 7377.2 & 7391.9 & 7425.7 & 7436.9
\\ \hline
Channels & $\bar{B}_s^0D^0$ & $B^-D_s^+$ & $\bar{B}_s^{*0}D^0$ & $B^{*-}D_s^+$ & $\bar{B}_s^0D^{*0}$ & $B^-D_s^{*+}$ & $\bar{B}_s^{*0}D^{*0}$ & $B^{*-}D_s^{*+}$
\\ \hline
Threshold & 7231.8 & 7247.7 & 7280.2 & 7293.1 & 7373.8 & 7391.5 & 7422.3 & 7436.9
\\ \hline
Channels & $\bar{B}^0D_s^-$ & $B_c^-\bar{K}^0$ & $\bar{B}^{*0}D_s^-$ & $B_c^{*-}\bar{K}^0$ & $\bar{B}^0D_s^{*-}$ & $B_c^-\bar{K}^{*0}$ & $\bar{B}^{*0}D_s^{*-}$ & $B_c^{*-}\bar{K}^{*0}$
\\ \hline
Threshold & 7247.7 & 6772.1 & 7293.1 & 6828.6 & 7391.9 & 7170.0 & 7436.9 & 7226.6
\\ \hline
Channels & $B^-D_s^-$ & $B_c^-K^-$ & $B^{*-}D_s^-$ & $B_c^{*-}K^-$ & $B^-D_s^{*-}$ & $B_c^-K^{*-}$ & $B^{*-}D_s^{*-}$ & $B_c^{*-}K^{*-}$
\\ \hline
Threshold & 7247.7 & 6768.2 & 7293.1 & 6824.7 & 7391.9 & 7166.1 & 7436.9 &  7222.7
\\ \hline
Channels & $\bar{B}_s^0D^-$ & $B_c^-K^0$ & $\bar{B}_s^{*0}D^-$ & $B_c^{*-}K^0$ & $\bar{B}_s^0D^{*-}$ & $B_c^-K^{*0}$ & $\bar{B}_s^{*0}D^{*-}$ & $B_c^{*-}K^{*0}$
\\ \hline
Threshold & 7236.6 & 6772.1 & 7285.1 & 6828.6 & 7377.2 & 7170.0 & 7425.7 & 7226.6
\\ \hline
Channels & $\bar{B}_s^0\bar{D}^0$ & $B_c^-K^+$ & $\bar{B}_s^{*0}\bar{D}^0$ & $B_c^{*-}K^+$ & $\bar{B}_s^0\bar{D}^{*0}$ & $B_c^-K^{*+}$ & $\bar{B}_s^{*0}\bar{D}^{*0}$ & $B_c^{*-}K^{*+}$
\\ \hline
Threshold & 7231.8 & 6768.2 & 7280.2 & 6824.7 & 7373.8 & 7166.1 & 7422.3 & 7222.7
\\ \hline\hline
\end{tabular}}
\label{tab:threshold}
\end{table*}

\begin{figure}[hbtp]
\centering
\includegraphics[width=0.95\linewidth]{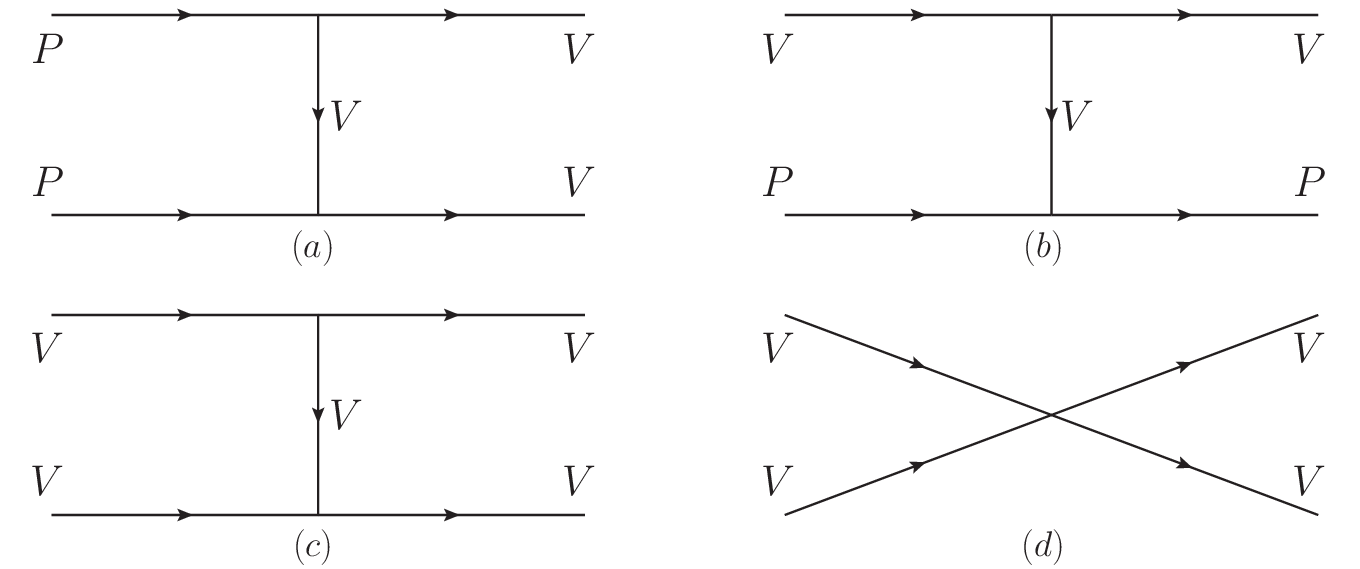}
\caption{Feynman diagrams for the interactions between charmed(-strange) mesons and bottom(-strange) mesons: (a) the vector meson exchange between two pseudoscalar mesons, (b) the vector meson exchange between vector and pseudoscalar mesons, (c) the vector meson exchange between two vector mesons, and (d) the contact term connecting four vector mesons. }
\label{fig-lagrangians}
\end{figure}

Within the extended local hidden gauge symmetry approach, the interactions between charmed(-strange) mesons and bottom(-strange) mesons mainly proceed through the exchange of the vector meson, as depicted in Figs.~\ref{fig-lagrangians}(a,b,c). Together with the contact term depicted in Fig.~\ref{fig-lagrangians}(d), their corresponding Lagrangians can be written as:
\begin{eqnarray}
\mathcal{L}_{VPP} &=& -ig \, \langle[P,\partial_{\mu} P] V^{\mu}\rangle \, ,
\label{eq-LVPP}
\\
\mathcal{L}_{VVV} &=& ig \, \langle(V^{\mu}\partial_{\nu}V_{\mu}-\partial_{\nu}V^{\mu}V_{\mu})V^{\nu}\rangle \, ,
\label{eq-LVVV}
\\
\mathcal{L}_{VVVV} &=& \frac{g^2}{2}\langle V_{\mu}V_{\nu}V^{\mu}V^{\nu}-V_{\nu}V_{\mu}V^{\mu}V^{\nu}\rangle \, .
\label{eq-LVVVV}
\end{eqnarray}
The coupling constant is defined as $g={M_V}/({2f_\pi})$, where $M_V$ is the mass of the exchanged vector meson and $f_\pi = 93$~MeV is the decay constant of pion. Especially, we shall take $M_V= 800 \ \textrm{MeV}$ for the mass of the exchanged light vector meson.

Taking into account the standard $\eta$-$\eta^{\prime}$ mixing, we can write the matrices of the flavor $SU(5)$ pseudoscalar and vector mesons as follows,
\begin{widetext}
\setlength{\arraycolsep}{2pt}
\renewcommand{\arraystretch}{2}
\begin{eqnarray}
P &=&
\left(\begin{array}{ccccc}
\frac{\eta}{\sqrt{3}}+\frac{\eta'}{\sqrt{6}}+\frac{\pi^0}{\sqrt{2}} & \pi^+                                                               & K^+                                            & ~~~~~~\bar{D}^0~~~~~~ & B^+   \\
\pi^-                                                               & \frac{\eta}{\sqrt{3}}+\frac{\eta'}{\sqrt{6}}-\frac{\pi^0}{\sqrt{2}} & K^{0}                                          & D^-                   & B^0   \\
K^{-}                                                               & \bar{K}^{0}                                                         & -\frac{\eta}{\sqrt{3}}+\sqrt{\frac{2}{3}}\eta' & D_s^-                 & B_s^0 \\
D^0 & D^+       & D_s^+       & \eta_c & B_c^+ \\
B^- & \bar{B}^0 & \bar{B}^0_s & B_c^-  & \eta_b
\end{array} \right)\, ,
\label{eq-pfields}
\\ V &=&
\left( \begin{array}{ccccc}
\frac{\omega+\rho^0}{\sqrt{2}} & \rho^+                         & K^{*+}         & \bar{D}^{*0} & B^{*+} \\
\rho^-                         & \frac{\omega-\rho^0}{\sqrt{2}} & K^{*0}         & D^{*-}       & B^{*0} \\
K^{*-}                         & \bar{K}^{*0}                   & \phi           & D_s^{*-}     & B_s^{*0} \\
~~~D^{*0}~~~                   & ~~~D^{*+}~~~                   & ~~~D_s^{*+}~~~ & ~~~J/\psi~~~ & ~~~B_c^{*+}~~~ \\
B^{*-}                         & \bar{B}^{*0}                   & \bar{B}^{*0}_s & B_c^{*-}     & \Upsilon
\end{array} \right) \, .
\label{eq-vfields}
\end{eqnarray}
\end{widetext}
Although the flavor $SU(5)$ symmetry has been used here, for the vector exchange between mesons which is the dominant part in the large quark mass counting, one is only using the $\bar q q$ character of the mesons~\cite{Sakai:2017avl}. Furthermore, in this work the heavy quarks in the coupled-channels are the bystanders according to the heavy quark symmetry, and the light quarks are the participants in the reactions, so the exchange of light vector mesons contributes dominantly to the $bc\bar{s}\bar{q}$ system. As discussed in Ref.~\cite{Debastiani:2017ewu}, the exchange of light vector mesons can be well described by the $SU(3)$ flavor symmetry. We shall further investigate the contribution from the exchange of heavy vector mesons in Appendix~\ref{sec:app} for the $b\bar{c}s\bar{q}$ and $b\bar{c}\bar{s}q$ systems, where the exchange of light vector mesons is not allowed.

We shall study the interaction of the $bc\bar{s}\bar{d}$ system as an example, and separately investigate the vector meson exchange between two pseudoscalar mesons, the vector meson exchange between vector and pseudoscalar mesons, and the vector meson exchange between two vector mesons in the following subsections. Studies on the $b\bar{c}s\bar{q}$ and $b\bar{c}\bar{s}q$ systems can be found in Appendix~\ref{sec:app}.

\subsection{$P$-$P$ interaction in the $bc\bar{s}\bar{d}$ system}

In this subsection we study the interaction due to the vector meson exchange between two pseudoscalar mesons in the $bc\bar{s}\bar{d}$ system, and the $bc\bar{s}\bar{u}$ system can be similarly investigated. There are only two coupled channels:
\begin{equation}
\nonumber
\bar{B}_s^0D^+ \, , \, \bar B^0D_s^+ \, .
\end{equation}

\begin{figure}[h]
\centering
\includegraphics[width=0.95\linewidth]{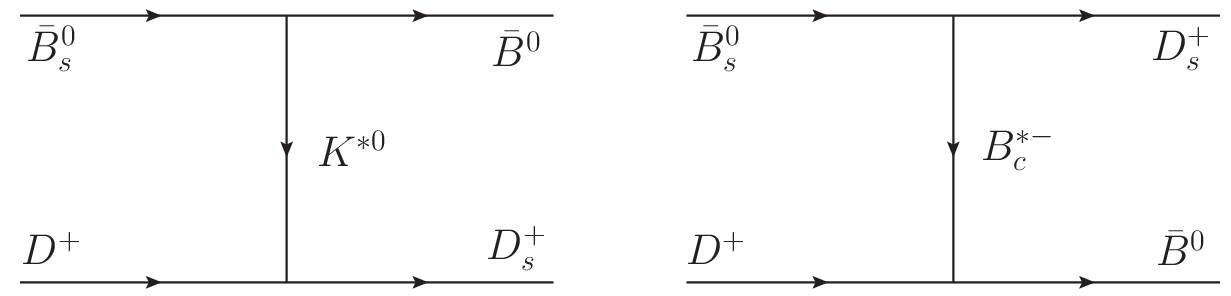}
\caption{The vector meson exchange between two pseudoscalar mesons in the $bc\bar{s}\bar{d}$ system.}
\label{fig-Bs0bD0interaction}
\end{figure}

As shown in Fig.~\ref{fig-Bs0bD0interaction}, the exchanged vector meson can be either the $K^{*0}$ or $B_c^{*-}$ meson. Since the latter $B_c^{*-}$ meson is too massive, we do not take it into account in this work.

Based on Eq.~(\ref{eq-LVPP}), the transition potential $V(s)$ due to the vector meson exchange between two pseudoscalar mesons can be written as
\begin{equation}
V_{PP}(s) = C_{PP} \times g^2 (p_1+p_3) (p_2+p_4) \, ,
\label{eq-Vij}
\end{equation}
with $p_1(p_3)$ the four-momentum of the $\bar{B}_s^0(\bar B^0)$ meson and $p_2(p_4)$ the four-momentum of the $D^+ (D_s^+)$ meson. The matrix $C_{PP}$ for the $bc\bar{s}\bar{d}$ system is a $2\times2$ matrix:
\begin{equation}
\setlength{\arraycolsep}{5pt}
\renewcommand{\arraystretch}{2}
C_{PP}=\left(
\begin{array}{c|cc}
J=0            & \bar{B}_s^0D^+      & \bar B^0D_s^+
\\ \hline
\bar{B}_s^0D^+ & 0                   & \frac{1}{m_{K^*}^2} \\
\bar B^0D_s^+       & \frac{1}{m_{K^*}^2} & 0
\end{array}
\right)\, ,
\label{eq-Cij}
\end{equation}
which does not contain any diagonal term.

As shown in Table~\ref{tab:threshold}, the threshold masses of the $\bar{B}_s^0D^+$ and $\bar B^0D_s^+$ channels are quite close to each other. This triggers us to consider their mixing:
\begin{eqnarray}
\nonumber |(\bar{B}D)_{s}^+; J=0 \rangle &=& \frac{1}{\sqrt{2}}\left(|\bar{B}_s^0D^+\rangle_{J=0} + |\bar B^0D_s^+\rangle_{J=0}\right),
\\ \label{def:BDs}
\\ \nonumber
|(\bar{B}D)_{s}^-; J=0 \rangle &=& \frac{1}{\sqrt{2}}\left(|\bar{B}_s^0D^+\rangle_{J=0} - |\bar B^0D_s^+\rangle_{J=0}\right),
\\
\end{eqnarray}
and the matrix $C_{PP}$ in this basis transforms to be
\begin{equation}
\setlength{\arraycolsep}{5pt}
\renewcommand{\arraystretch}{2}
C_{PP}^\prime=\left(
\begin{array}{c|cc}
J=0                & (\bar{B}D)_{s}^+    & (\bar{B}D)_{s}^-
\\ \hline
(\bar{B}D)_{s}^+   & \frac{1}{m_{K^*}^2} & 0 \\
(\bar{B}D)_{s}^-   & 0                   & -\frac{1}{m_{K^*}^2}
\end{array}
\right)\, .
\label{eq-Cijr}
\end{equation}
Accordingly, the attractive combination $|(\bar{B}D)_{s}^-; J=0 \rangle$ may produce a bound state, while the repulsive combination $|(\bar{B}D)_{s}^+; J=0 \rangle$ can not.

\subsection{$V$-$P$ interaction in the $bc\bar{s}\bar{d}$ system}

\begin{figure}[h]
\centering
\includegraphics[width=.9\linewidth]{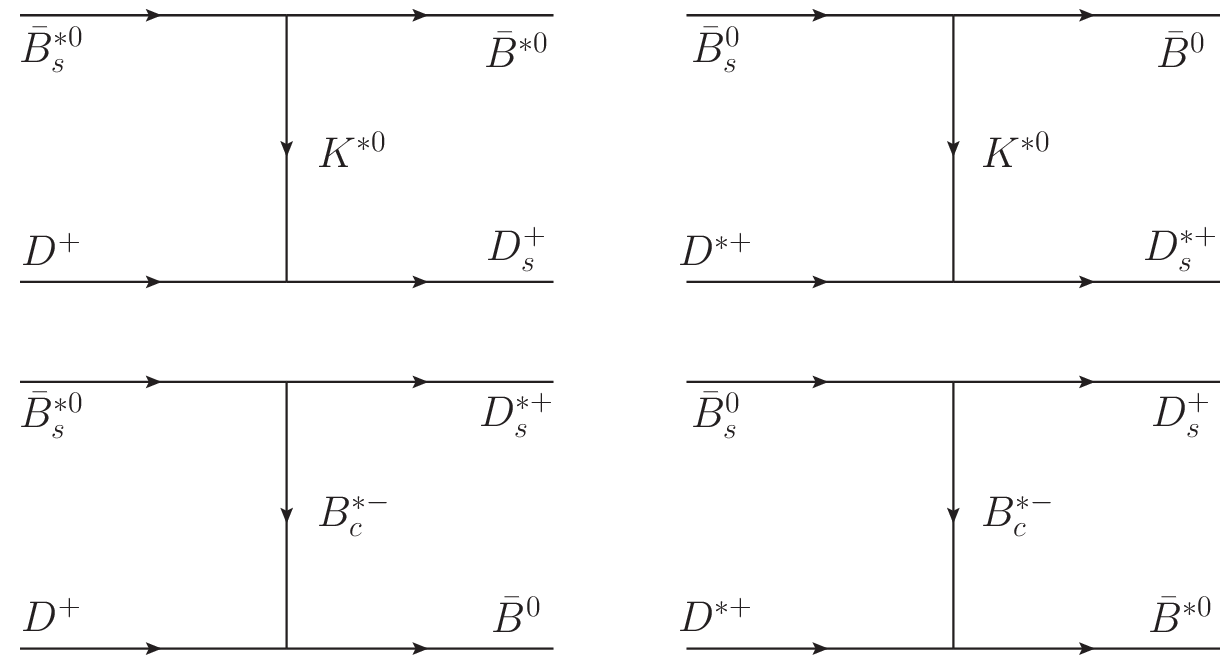}
\caption{The vector meson exchange between vector and pseudoscalar mesons in the $bc\bar{s}\bar{d}$ system.}
\label{fig-Bsst0bD0interaction}
\end{figure}

In this subsection we study the interaction due to the vector meson exchange between vector and pseudoscalar mesons in the $bc\bar{s}\bar{d}$ system. There are four coupled channels:
\begin{equation}
\nonumber
\bar{B}_s^{*0}D^+ \, , \, \bar B^{*0}D_s^+ \, , \, \bar{B}_s^0D^{*+} \, , \, \bar B^{*0} D_s^{*+} \, .
\end{equation}
As shown in Fig.~\ref{fig-Bsst0bD0interaction}, the exchanged vector meson can still be either the $K^{*0}$ or $B_c^{*-}$ meson, and again we neglect the latter due to its large mass.

Based on Eq.~(\ref{eq-LVPP}) and Eq.~(\ref{eq-LVVV}), the transition potential $V(s)$ due to the vector meson exchange between vector and pseudoscalar mesons can be written as
\begin{equation}
V_{VP}(s) = C_{VP} \times g^2 (p_1+p_3) (p_2+p_4) ~ \vec\epsilon \cdot \vec\epsilon\,' \, ,
\label{eq-Vij-VV-K}
\end{equation}
with the $4\times4$ matrix
\begin{eqnarray}
&&C_{VP} =
\\ \nonumber &&
\renewcommand{\arraystretch}{1.5}
\left(
\begin{array}{c|cccc}
J=1 & \bar{B}_s^{*0}D^+ & \bar B^{*0}D_s^+ & \bar{B}_s^0D^{*+} & \bar B^0 D_s^{*+}
\\ \hline
\bar{B}_s^{*0}D^+ & 0 & \frac{1}{m_{K^*}^2} & 0 & 0 \\
\bar B^{*0}D_s^+       & \frac{1}{m_{K^*}^2} & 0 & 0 & 0 \\
\bar{B}_s^0D^{*+} & 0 & 0 &  0 & \frac{1}{m_{K^*}^2} \\
\bar B^0 D_s^{*+}      & 0 & 0 & \frac{1}{m_{K^*}^2} &  0
\end{array}
\right)\, .
\label{eq-Cijr-VP}
\end{eqnarray}

Since the three-momenta of the external vector mesons can be ignored compared to their masses when working at the threshold, we approximate $\epsilon^0 \approx 0$ for these external vector mesons. This makes the $VP \to VP$ transition potential similar to the $PP \to PP$ transition potential, and we just need to add an extra factor $\vec\epsilon \cdot \vec\epsilon\,'$, with $\vec\epsilon(\vec\epsilon\,')$ the polarization vector of the initial(final) vector meson.

As shown in Table~\ref{tab:threshold}, the threshold masses of the $\bar{B}_s^{*0}D^+$ and $\bar B^{*0}D_s^+$ channels are close to each other, and the threshold masses of the $\bar{B}_s^0D^{*+}$ and $\bar B^0 D_s^{*+}$ channels are also close to each other. Accordingly, we consider the four mixed channels:
\begin{eqnarray}
\nonumber |(\bar{B}^*D)_{s}^+; J=1 \rangle &=& \frac{1}{\sqrt{2}}\left(|\bar{B}_s^{*0}D^+\rangle_{J=1} + |\bar B^{*0}D_s^+\rangle_{J=1}\right),
\\
\\ \nonumber
|(\bar{B}^*D)_{s}^-; J=1 \rangle &=& \frac{1}{\sqrt{2}}\left(|\bar{B}_s^{*0}D^+\rangle_{J=1} - |\bar B^{*0}D_s^+\rangle_{J=1}\right),
\\
\\ \nonumber
|(\bar{B}D^*)_{s}^+; J=1 \rangle &=& \frac{1}{\sqrt{2}}\left(|\bar{B}_s^0D^{*+}\rangle_{J=1} + |\bar B^0D_s^{*+}\rangle_{J=1}\right),
\\
\\ \nonumber
|(\bar{B}D^*)_{s}^-; J=1 \rangle &=& \frac{1}{\sqrt{2}}\left(|\bar{B}_s^0D^{*+}\rangle_{J=1} - |\bar B^0D_s^{*+}\rangle_{J=1}\right),
\\
\end{eqnarray}
and the matrix $C_{VP}$ in this basis transforms to be
\begin{eqnarray}
&& C_{VP}^\prime =
\\ \nonumber &&
\renewcommand{\arraystretch}{1.5}
\left(
\begin{array}{c|cccc}
J=1 & (\bar{B}^*D)_{s}^+ & (\bar{B}^*D)_{s}^- & (\bar{B}D^*)_{s}^+ & (\bar{B}D^*)_{s}^-
\\ \hline
(\bar{B}^*D)_{s}^+ & \frac{1}{m_{K^*}^2} & 0 & 0 & 0 \\
(\bar{B}^*D)_{s}^- & 0 & -\frac{1}{m_{K^*}^2} & 0 & 0 \\
(\bar{B}D^*)_{s}^+ & 0 & 0 & \frac{1}{m_{K^*}^2} & 0 \\
(\bar{B}D^*)_{s}^- & 0 & 0 & 0 & -\frac{1}{m_{K^*}^2}
\end{array}
\right)\, .
\end{eqnarray}
Accordingly, the attractive combinations $|(\bar{B}^*D)_{s}^-; J=1 \rangle$ and $|(\bar{B}D^*)_{s}^-; J=1 \rangle$ may produce two bound states, while the repulsive combinations $|(\bar{B}^*D)_{s}^+; J=1 \rangle$ and $|(\bar{B}D^*)_{s}^+; J=1 \rangle$ can not.

\subsection{$V$-$V$ interaction in the $bc\bar{s}\bar{d}$ system}

\begin{figure}[h]
\centering
\includegraphics[width=.9\linewidth]{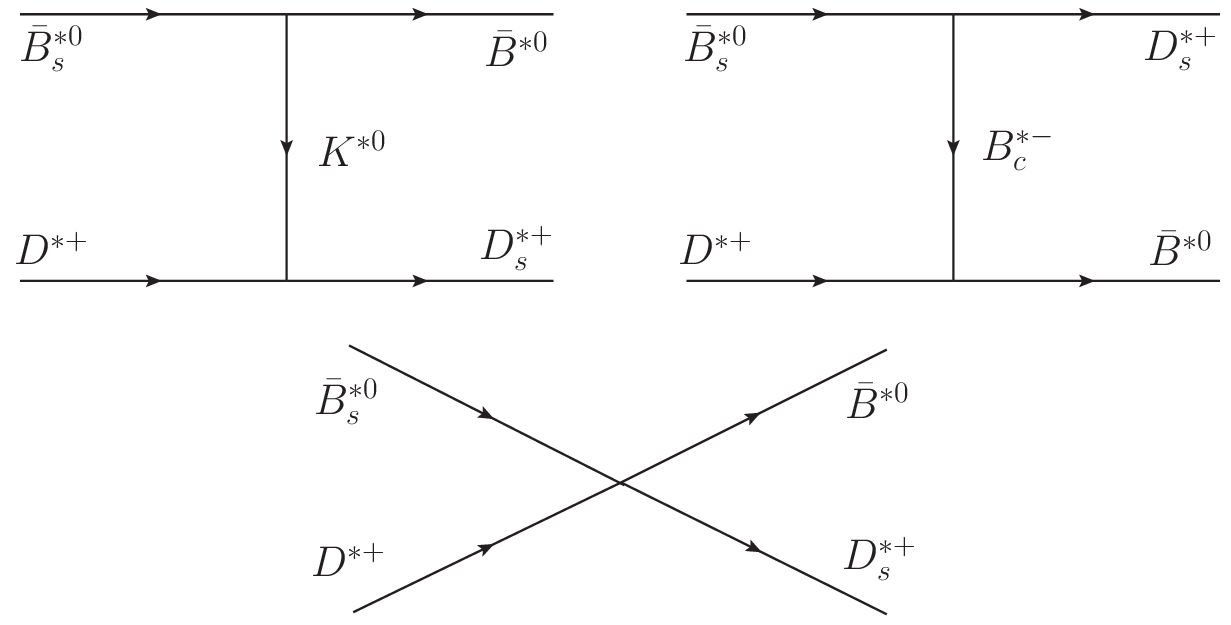}
\caption{The vector meson exchange between two vector mesons in the $bc\bar{s}\bar{d}$ system as well as the contact term connecting four vector mesons.}
\label{fig-VV-intercation}
\end{figure}

 In this subsection we study the interaction due to the vector meson exchange between two vector mesons in the $bc\bar{s}\bar{d}$ system. There are two coupled channels:
\begin{equation}
\nonumber
\bar{B}_s^{*0}D^{*+} \, , \, \bar B^{*0}D_s^{*+} \, .
\end{equation}
As shown in Fig.~\ref{fig-VV-intercation}, the exchanged vector meson can be either the $K^{*0}$ or $B_c^{*-}$ meson, and we again neglect the contribution from the $B_c^{*-}$ meson exchange due to its large mass. Besides, we also need to take into account the contact term connecting four vector mesons, so the complete $VV$ transition potential consists of two terms:
\begin{equation}
V_{VV}(s)=V_{VV}(s)^{ex}+V_{VV}(s)^{co} \, .
\label{eq:Vij-VV-total}
\end{equation}

Based on Eq.~(\ref{eq-LVVV}), the transition potential $V_{VV}(s)^{ex}$ due to the vector meson exchange between two vector mesons can be written as
\begin{equation}
V_{VV}(s)^{ex} = C_{VV} \times g^2 (p_1+p_3) (p_2+p_4)\epsilon_1 \cdot \epsilon_3 \epsilon_2 \cdot \epsilon_4 \, ,
\label{eq-Vij-VV-ex}
\end{equation}
where $p_1(p_3)$ is the four-momentum of the $\bar{B}_s^{*0}(\bar B^{*0})$ meson, $p_2(p_4)$ is the four-momentum of the $D^{*+} (D_s^{*+})$ meson, $\epsilon_1(\epsilon_3)$ is the polarization vector of the $\bar{B}_s^{*0}(\bar B^{*0})$ meson, and $\epsilon_2(\epsilon_4)$ is the polarization vector of the $D^{*+} (D_s^{*+})$ meson. The matrix $C_{VV}$ for the $bc\bar{s}\bar{d}$ system is a $2\times2$ matrix:
\begin{equation}
\setlength{\arraycolsep}{5pt}
\renewcommand{\arraystretch}{2}
C_{VV}=\left(
\begin{array}{c|cc}
J=0,1,2            & \bar{B}_s^{*0}D^{*+}      & \bar B^{*0}D_s^{*+}
\\ \hline
\bar{B}_s^{*0}D^{*+} & 0                   & \frac{1}{m_{K^*}^2} \\
\bar B^{*0}D_s^{*+}       & \frac{1}{m_{K^*}^2} & 0
\end{array}
\right)\, ,
\label{eq-Cij}
\end{equation}
which does not contain any diagonal term.

In addition, the transition potential $V_{VV}(s)^{co}$ can be extracted from Eq.~(\ref{eq-LVVVV}) to be
\begin{equation}
\begin{split}
V_{VV}(s)^{co} &=
m_{K^*}^2 \cdot  C_{VV} \\
&\times g^2(-2\epsilon_\mu\epsilon^\mu\epsilon_\nu\epsilon^\nu + \epsilon_\mu\epsilon_\nu\epsilon^\mu\epsilon^\nu + \epsilon_\mu\epsilon_\nu\epsilon^\nu\epsilon^\mu) \, .
\label{V-contact}
\end{split}
\end{equation}
By using the spin projection operators,
\begin{eqnarray}
{\cal P}^{(0)}&=& \frac{1}{3}\epsilon_\mu \epsilon^\mu \epsilon_\nu \epsilon^\nu\nonumber\\
{\cal P}^{(1)}&=&\frac{1}{2}(\epsilon_\mu\epsilon_\nu\epsilon^\mu\epsilon^\nu-\epsilon_\mu\epsilon_\nu\epsilon^\nu\epsilon^\mu)\nonumber\\
{\cal P}^{(2)}&=&\frac{1}{2}(\epsilon_\mu\epsilon_\nu\epsilon^\mu\epsilon^\nu+\epsilon_\mu\epsilon_\nu\epsilon^\nu\epsilon^\mu)-\frac{1}{3}\epsilon_\mu\epsilon^\mu\epsilon_\nu\epsilon^\nu\ ,
\label{eq:projmu}
\end{eqnarray}
Eq.~(\ref{V-contact}) can be written separately for the spin $J=0/1/2$ channels as:
\begin{eqnarray}
V_{VV}(s)^{co}= m_{K^*}^2 \cdot C_{VV} \times
\left\{\begin{array}{cc}
-4 g^2&~\textrm{for $J=0$},\\
0  &~\textrm{for $J=1$},\\
2 g^2&~\textrm{for $J=2$}.
\end{array}\right.
\label{eq-contact-in-different-J}
\end{eqnarray}

As shown in Table \ref{tab:threshold}, the threshold masses of $\bar{B}_s^{*0}D^{*+}$ and $\bar B^{*0}D_s^{*+}$ channels are close to each other, so we consider the two mixed channels as we did in the previous subsections:
\begin{eqnarray}
&& |(\bar{B}^*D^*)_{s}^+; J=0,1,2 \rangle
\\ \nonumber && ~~~~~~~~~~ = \frac{1}{\sqrt{2}}\left(|\bar{B}_s^{*0}D^{*+}\rangle_{J=0,1,2} + |\bar B^{*0}D_s^{*+}\rangle_{J=0,1,2}\right),
\\ && |(\bar{B}^*D^*)_{s}^-; J=0,1,2 \rangle
\\ \nonumber && ~~~~~~~~~~ = \frac{1}{\sqrt{2}}\left(|\bar{B}_s^{*0}D^{*+}\rangle_{J=0,1,2} - |\bar B^{*0}D_s^{*+}\rangle_{J=0,1,2}\right).
\end{eqnarray}
The matrix $C_{VV}$ in this basis transforms to be
\begin{equation}
\setlength{\arraycolsep}{5pt}
\renewcommand{\arraystretch}{2}
C_{VV}^\prime=\left(
\begin{array}{c|cc}
J=0,1,2                & (\bar{B}^*D^*)_{s}^+    & (\bar{B}^*D^*)_{s}^-
\\ \hline
(\bar{B}^*D^*)_{s}^+   & \frac{1}{m_{K^*}^2} & 0 \\
(\bar{B}^*D^*)_{s}^-   & 0                   & -\frac{1}{m_{K^*}^2}
\end{array}
\right)\, .
\label{eq-Cijr-VV}
\end{equation}

It is worth mentioning that the contribution of the contact term $V_{VV}(s)^{co}$, subleading in the heavy quark mass counting, is much smaller than the vector meson exchange term $V_{VV}(s)^{ex}$, so the combinations $|(\bar{B}^*D^*)_{s}^-; J=0,1,2 \rangle$ are still attractive, while the combinations $|(\bar{B}^*D^*)_{s}^+; J=0,1,2 \rangle$ are always repulsive.

\section{Numerical Results}
\label{sec:result}

In this section we perform numerical analyses to study the hadronic molecular states with the quark contents $bc\bar{s}\bar{q}$, $b\bar{c}s\bar{q}$, and $b\bar{c}\bar{s}q$. Again we use the $bc\bar{s}\bar{d}$ system as an example. Based on Eq.~(\ref{eq-Vij}), Eq.~(\ref{eq-Vij-VV-K}), Eq.~(\ref{eq:Vij-VV-total}), Eq.~(\ref{eq-Vij-VV-ex}), and Eq.~(\ref{eq-contact-in-different-J}), we can solve the Bethe-Salpeter equation to obtain the scattering amplitude
\begin{equation}
T_{PP/VP/VV}(s) = {V_{PP/VP/VV}(s) \over 1 - V_{PP/VP/VV}(s) G(s)} \, ,
\label{eq-bseq}
\end{equation}
where $G(s)$ is the diagonal loop function
\begin{equation}
G_{ii}(s) = i\int{\frac{d^4q}{(2\pi)^4}\frac{1}{q^2-m_1^2+i\epsilon}\frac{1}{(p-q)^2-m_2^2+i\epsilon}} \, .
\label{eq-G}
\end{equation}
Here $s = p^2$ with $p$ the total four-momentum; $m_1$ and $m_2$ are the masses of the two mesons involved in the present channel.

We regularize Eq.~(\ref{eq-G}) through the cut-off method as
\begin{equation}
G_{ii}(s) = \int_0^{q_\textrm{max}} \frac{d^3q}{(2\pi)^3} \frac{\omega_1 + \omega_2}{2\omega_1\omega_2} \frac{1}{s-(\omega_1+\omega_2)^2+i\epsilon} \, ,
\label{eq-cutoff-G}
\end{equation}
with $\omega_1 = \sqrt{m_1^2+\vec{q}^{\,2}}$ and $\omega_2 = \sqrt{m_2^2+\vec{q}^{\,2}}$. In this work we take two values for the cut-off momentum, $q_\textrm{max} = 400$~MeV and $q_\textrm{max} = 600$~MeV, meanwhile the similar cut-off momentum $q_{\textrm{max}}$ is used to study the meson-meson interaction with heavy flavors in Refs.~\cite{Feijoo:2021ppq,Sakai:2017avl,Dai:2022ulk}.

Equation.~(\ref{eq-cutoff-G}) holds on the physical sheet, {\it i.e.}, the first Riemann sheet. Sometimes we also need to search for poles on the second Riemann sheet. In the latter case we define $G_{ii}^{II}(s)$  as
\begin{equation}
G_{ii}^{II}(s) = G_{ii}(s) + i {k \over 4 \pi \sqrt s} \, ,
\end{equation}
with $k(s) = \sqrt{ (s-(m_1+m_2)^2)(s-(m_1-m_2)^2) } / (2\sqrt s) $ for Im$(k)>0$.

In order to express the coupling strength of the pole to different channels, we introduce the coupling $g_i$ and define it in the vicinity of the pole as
\begin{equation}
\label{eq:gi1}
T_{ij}(s) = \frac{g_ig_j}{s-s_p^2} \, .
\end{equation}
Here $s_p$ is the position of pole on the $\sqrt{s}$ complex plane, and the coupling $g_i$ is the coupling constant between the pole and the channel $i$. We can also write it in the residue form as
\begin{equation}
\label{eq:gi2}
g_i^2 = \lim_{\sqrt{s} \to s_p} (s-s_p^2)~T_{ii}(s) \, .
\end{equation}

\begin{table}[htb]
\centering
\renewcommand{\arraystretch}{1.5}
\caption{The binding energies $E_B$ and the couplings $g_i$ of the bound states on the physical (first Riemann) sheet for the $bc\bar{s}\bar{d}$ and $bc\bar{s}\bar{u}$ systems, with the cut-off momentum $q_\textrm{max}=600\mev$. The binding energies for the $bc\bar{s}\bar{d}$ system are relevant to the lower $\bar{B}_s^{(*)0}D^{(*)+}$ channels other than the higher $\bar{B}^{(*)0}D_s^{(*)+}$ channels, and the binding energies for the $bc\bar{s}\bar{u}$ system are relevant to the lower $\bar{B}_s^{(*)0}D^{(*)0}$ channels other than the higher $B^{(*)-}D_s^{(*)+}$ channels .}
\begin{tabular}{c|c|c|c|c}
\hline\hline
Content: $bc\bar{s}\bar{d}$ & \,$I(J^P)$\, &$E_B$ (MeV) & \,Channel\, & $|g_i|$ (GeV)
\\ \hline\hline
\multirow{2}{*}{$|(\bar{B}D)_{s}^-; J=0 \rangle$} & \multirow{2}{*}{$\frac{1}{2}(0^+)$} & \multirow{2}{*}{15.7} & $\bar{B}_s^0D^+$        & 19
\\ \cline{4-5}
                                                                                                              &&& $\bar{B}^0D_s^+$        & 21  	
\\ \hline
\multirow{2}{*}{$|(\bar{B}^*D)_{s}^-; J=1 \rangle$} & \multirow{2}{*}{$\frac{1}{2}(1^+)$} & \multirow{2}{*}{17.3}  & $\bar{B}_s^{*0}D^+$  & 20
\\ \cline{4-5}
                                                                                                                 &&& $\bar{B}^{*0}D_s^+$  & 21  	
\\ \hline
\multirow{2}{*}{$|(\bar{B}D^*)_{s}^-; J=1 \rangle$} & \multirow{2}{*}{$\frac{1}{2}(1^+)$} & \multirow{2}{*}{16.4}  & $\bar{B}_s^0D^{*+}$  & 20
\\ \cline{4-5}
                                                                                                                 &&& $\bar{B}^0D_s^{*+}$  & 23  	

\\ \hline
\multirow{2}{*}{$|(\bar{B}^*D^*)_{s}^-; J=0 \rangle$} & \multirow{2}{*}{$\frac{1}{2}(0^+)$} & \multirow{2}{*}{13.6} & $\bar{B}_s^{*0}D^{*+}$ & 19
\\ \cline{4-5}
                                                                                                                  &&& $\bar B^{*0}D_s^{*+}$       & 21
\\ \hline
\multirow{2}{*}{$|(\bar{B}^*D^*)_{s}^-; J=1 \rangle$} & \multirow{2}{*}{$\frac{1}{2}(1^+)$} & \multirow{2}{*}{18.2} & $\bar{B}_s^{*0}D^{*+}$ & 21
\\ \cline{4-5}
                                                                                                                  &&& $\bar B^{*0}D_s^{*+}$       & 23
\\ \hline
\multirow{2}{*}{$|(\bar{B}^*D^*)_{s}^-; J=2 \rangle$} & \multirow{2}{*}{$\frac{1}{2}(2^+)$} & \multirow{2}{*}{20.5} & $\bar{B}_s^{*0}D^{*+}$ & 22
\\ \cline{4-5}
                                                                                                                  &&& $\bar B^{*0}D_s^{*+}$       & 24
\\ \hline\hline

Content: $bc\bar{s}\bar{u}$ & \,$I(J^P)$\, &$E_B$ (MeV) & \,Channel\, & $|g_i|$ (GeV)
\\ \hline\hline
\multirow{2}{*}{$|(\bar{B}D)_{s}^-; J=0 \rangle^\prime$} & \multirow{2}{*}{$\frac{1}{2}(0^+)$} & \multirow{2}{*}{14.3} & $\bar{B}_s^0D^0$ & 19
\\ \cline{4-5}
                                                                                                              &&& $B^-D_s^+$       & 22  	
\\ \hline
\multirow{2}{*}{$|(\bar{B}^*D)_{s}^-; J=1 \rangle^\prime$} & \multirow{2}{*}{$\frac{1}{2}(1^+)$} & \multirow{2}{*}{15.7}  & $\bar{B}_s^{*0}D^0$  & 19
\\ \cline{4-5}
                                                                                                                 &&& $B^{*-}D_s^+$        & 22  	
\\ \hline
\multirow{2}{*}{$|(\bar{B}D^*)_{s}^-; J=1 \rangle^\prime$} & \multirow{2}{*}{$\frac{1}{2}(1^+)$} & \multirow{2}{*}{15.8}  & $\bar{B}_s^0D^{*0}$  & 20
\\ \cline{4-5}
                                                                                                                 &&& $B^-D_s^{*+}$        & 23  	

\\ \hline
\multirow{2}{*}{$|(\bar{B}^*D^*)_{s}^-; J=0 \rangle^\prime$} & \multirow{2}{*}{$\frac{1}{2}(0^+)$} & \multirow{2}{*}{12.7} & $\bar{B}_s^{*0}D^{*0}$ & 19
\\ \cline{4-5}
                                                                                                                  &&& $B^{*-}D_s^{*+}$       & 21
\\ \hline
\multirow{2}{*}{$|(\bar{B}^*D^*)_{s}^-; J=1 \rangle^\prime$} & \multirow{2}{*}{$\frac{1}{2}(1^+)$} & \multirow{2}{*}{17.2} & $\bar{B}_s^{*0}D^{*0}$ & 21
\\ \cline{4-5}
                                                                                                                  &&& $B^{*-}D_s^{*+}$       & 23
\\ \hline
\multirow{2}{*}{$|(\bar{B}^*D^*)_{s}^-; J=2 \rangle^\prime$} & \multirow{2}{*}{$\frac{1}{2}(2^+)$} & \multirow{2}{*}{19.5} & $\bar{B}_s^{*0}D^{*0}$ & 22
\\ \cline{4-5}
                                                                                                                  &&& $B^{*-}D_s^{*+}$       & 24
\\ \hline\hline

\end{tabular}
\label{tab:results1}
\end{table}

Firstly, we use the cut-off momentum $q_\textrm{max}=600\mev$ to perform numerical analyses. We find altogether six bound states in the $bc\bar{s}\bar{d}$ system with the binding energies about $10$-$20$~MeV: one state of $J^P=0^+$ below the $\bar{B}_s^0D^+$-$\bar B^0D_s^+$ threshold, one state of $J^P=1^+$ below the $\bar{B}_s^{*0}D^+$-$\bar B^{*0}D_s^+$ threshold, one state of $J^P=1^+$ below the $\bar{B}_s^0D^{*+}$-$\bar B^0D_s^{*+}$ threshold, and three states of $J^P=0^+/1^+/2^+$ below the $\bar{B}_s^{*0}D^{*+}$-$\bar B^{*0}D_s^{*+}$ threshold. We summarize their results in Table~\ref{tab:results1}. Besides, we have investigated the $bc\bar{s}\bar{u}$ system, where we also find six bound states. Similar to Eq.~(\ref{def:BDs}), we denote them as $|(\bar{B}^{(*)}D^{(*)})_{s}^-; J \rangle^\prime$, and summarize their results also in Table~\ref{tab:results1}. The parameter $|g_i|$ is about 20~GeV for all the channels, indicating that the mixing is roughly balanced, {\it e.g.}, the couplings of the mixing state $|(\bar{B}D)_{s}^-; J=0 \rangle$ to both the $\bar{B}_s^0D^+$ and $\bar B^0D_s^+$ channels are roughly equivalent.

Note that all these bound states have zero width. This is partly because: a) we do not consider the widths of the initial and final states, and b) we do not consider the box diagrams with pion exchanges, {\it e.g.}, see Ref.~\cite{Dai:2022ulk,Oset:2022xji} for discussions on these diagrams. Besides, we have neglected the exchange of the $B_c^*$ meson, so the bound state $|(\bar{B}D^*)_{s}^-; J=1 \rangle$ located at 7361~MeV only couples to the $\bar{B}_s^0D^{*+}$ and $\bar B^0D_s^{*+}$ channels, while it does not couple to the $\bar{B}_s^{*0}D^{+}$ and $\bar B^{*0}D_s^{+}$ channels, even if it lies above the thresholds of the $\bar{B}_s^{*0}D^{+}$ and $\bar B^{*0}D_s^{+}$ channels.

\begin{table}[htb]
\centering
\renewcommand{\arraystretch}{1.5}
\caption{The pole positions on the second Riemann sheet for the $bc\bar{s}\bar{d}$ and $bc\bar{s}\bar{u}$ systems as well as their corresponding threshold masses, with the cut-off momentum $q_\textrm{max}=400\mev$. Pole positions and threshold masses are both in units of MeV.}
\begin{tabular}{c|c|c|c|c}
\hline\hline
Content: $bc\bar{s}\bar{d}$ & $I(J^P)$ & Pole & \,Channel\, & Threshold
\\ \hline\hline
\multirow{2}{*}{$|(\bar{B}D)_{s}^-; J=0 \rangle$} & \multirow{2}{*}{$\frac{1}{2}(0^+)$} & \multirow{2}{*}{7235.2+$i$0} & $\bar{B}_s^0D^+$        & 7236.6
\\ \cline{4-5}
                                                                                                                            &&& $\bar{B}^0D_s^+$        & 7248.0
\\ \hline
\multirow{2}{*}{$|(\bar{B}^*D)_{s}^-; J=1 \rangle$} & \multirow{2}{*}{$\frac{1}{2}(1^+)$} & \multirow{2}{*}{7284.9+$i$0}  & $\bar{B}_s^{*0}D^+$  & 7285.1
\\ \cline{4-5}
                                                                                                                               &&& $\bar{B}^{*0}D_s^+$  & 7293.1
\\ \hline
\multirow{2}{*}{$|(\bar{B}D^*)_{s}^-; J=1 \rangle$} & \multirow{2}{*}{$\frac{1}{2}(1^+)$} & \multirow{2}{*}{7375.7+$i$0}  & $\bar{B}_s^0D^{*+}$  & 7377.2
\\ \cline{4-5}
                                                                                                                               &&& $\bar{B}^0D_s^{*+}$  &	7391.9

\\ \hline
\multirow{2}{*}{$|(\bar{B}^*D^*)_{s}^-; J=0 \rangle$} & \multirow{2}{*}{$\frac{1}{2}(0^+)$} & \multirow{2}{*}{7423.0+$i$0}  & $\bar{B}_s^{*0}D^{*+}$ & 7425.7
\\ \cline{4-5}
                                                                                                                                 &&& $\bar B^{*0}D_s^{*+}$       & 7436.9
\\ \hline
\multirow{2}{*}{$|(\bar{B}^*D^*)_{s}^-; J=1 \rangle$} & \multirow{2}{*}{$\frac{1}{2}(1^+)$} & \multirow{2}{*}{7425.4+$i$0}  & $\bar{B}_s^{*0}D^{*+}$ & 7425.7
\\ \cline{4-5}
                                                                                                                                 &&& $\bar B^{*0}D_s^{*+}$       & 7436.9
\\ \hline
\multirow{2}{*}{$|(\bar{B}^*D^*)_{s}^-; J=2 \rangle$} & \multirow{2}{*}{$\frac{1}{2}(2^+)$} & \multirow{2}{*}{7425.6+$i$0}  & $\bar{B}_s^{*0}D^{*+}$ & 7425.7
\\ \cline{4-5}
                                                                                                                                 &&& $\bar B^{*0}D_s^{*+}$       & 7436.9
\\ \hline\hline

Content: $bc\bar{s}\bar{u}$ & $I(J^P)$ & ~~~~Pole~~~~ & \,Channel\, & \,Threshold\,
\\ \hline
\multirow{2}{*}{$|(\bar{B}D)_{s}^-; J=0 \rangle^\prime$}  & \multirow{2}{*}{$\frac{1}{2}(0^+)$} & \multirow{2}{*}{7226.7+$i$0}   & $\bar{B}_s^0D^0$     &	7231.8
\\ \cline{4-5}
                                                                                                                        &&& $B^-D_s^+$           & 7247.7
\\ \hline
\multirow{2}{*}{$|(\bar{B}^*D)_{s}^-; J=1 \rangle^\prime$} & \multirow{2}{*}{$\frac{1}{2}(1^+)$} & \multirow{2}{*}{7278.4+$i$0}  & $\bar{B}_s^{*0}D^0$  & 7280.2
\\ \cline{4-5}
                                                                                                                        &&& $B^{*-}D_s^+$        & 7293.1
\\ \hline
\multirow{2}{*}{$|(\bar{B}D^*)_{s}^-; J=1 \rangle^\prime$} & \multirow{2}{*}{$\frac{1}{2}(1^+)$} & \multirow{2}{*}{7370.8+$i$0}  & $\bar{B}_s^0D^{*0}$  & 7373.8
\\ \cline{4-5}
                                                                                                                        &&& $B^-D_s^{*+}$        & 7391.5

\\ \hline
\multirow{2}{*}{$|(\bar{B}^*D^*)_{s}^-; J=0 \rangle^\prime$} & \multirow{2}{*}{$\frac{1}{2}(0^+)$} & \multirow{2}{*}{7415.6+$i$0} & $\bar{B}_s^{*0}D^{*0}$ & 7422.3
\\ \cline{4-5}
                                                                                                                         &&& $B^{*-}D_s^{*+}$       & 7436.9
\\ \hline
\multirow{2}{*}{$|(\bar{B}^*D^*)_{s}^-; J=1 \rangle^\prime$} & \multirow{2}{*}{$\frac{1}{2}(1^+)$} & \multirow{2}{*}{7421.2+$i$0} & $\bar{B}_s^{*0}D^{*0}$ & 7422.3
\\ \cline{4-5}
                                                                                                                         &&& $B^{*-}D_s^{*+}$       & 7436.9
\\ \hline
\multirow{2}{*}{$|(\bar{B}^*D^*)_{s}^-; J=2 \rangle^\prime$} & \multirow{2}{*}{$\frac{1}{2}(2^+)$} & \multirow{2}{*}{7421.9+$i$0} & $\bar{B}_s^{*0}D^{*0}$ & 7422.3
\\ \cline{4-5}
                                                                                                                         &&& $B^{*-}D_s^{*+}$       & 7436.9
\\ \hline\hline

\end{tabular}
\label{tab:results2}
\end{table}

Secondly, we use the cut-off momentum $q_\textrm{max}=400\mev$ to perform numerical analyses. In this case we do not find any bound state, {\it i.e.}, we do not find any pole in the first Riemann sheet below their corresponding thresholds. However, we find on the second Riemann sheet six poles for the $bc\bar{s}\bar{d}$ system, and the other six poles for the $bc\bar{s}\bar{u}$ system. As summarized in Table~\ref{tab:results2}, all these poles are below their corresponding thresholds, indicating their nature as the near-threshold virtual states.

\begin{figure}[h]
\centering
\includegraphics[width=.9\linewidth]{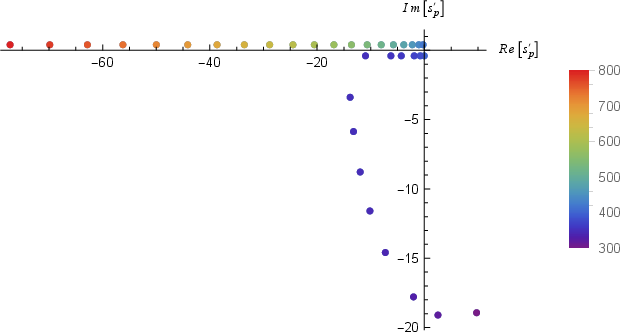}
\caption{The pole position $s_p^\prime = s_p - M_{\bar{B}_s^{*0}} - M_{D^{*+}}$ of the combination $|(\bar{B}^*D^*)_{s}^-; J=2 \rangle$ as a function of the cut-off momentum $q_{\textrm{max}} = 300 \sim 800$~MeV.}
\label{fig:polepostion1}
\end{figure}

\begin{figure*}[hbtp]
\centering
\subfigbottomskip=2pt
\subfigcapskip=-5pt
\subfigure[$\, |t|^2$ with $q_{\textrm{max}}=700$ ]{
\includegraphics[width=0.3\linewidth]{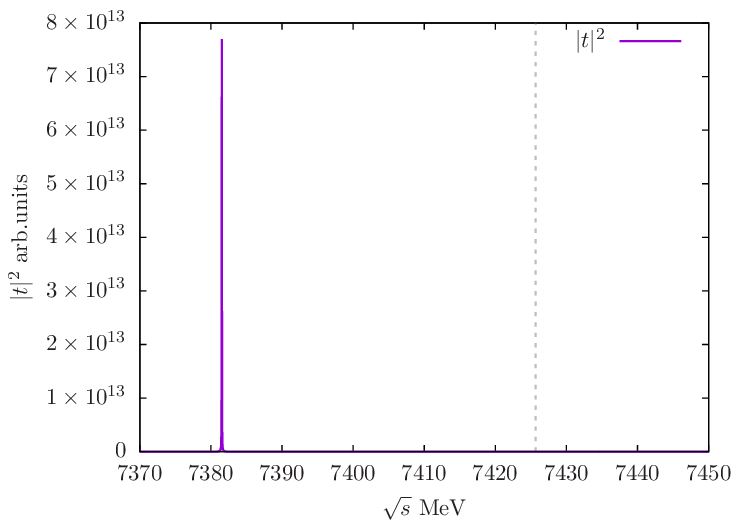}}
\subfigure[$\, |t|^2$ with $q_{\textrm{max}}=600$ ]{
\includegraphics[width=0.3\linewidth]{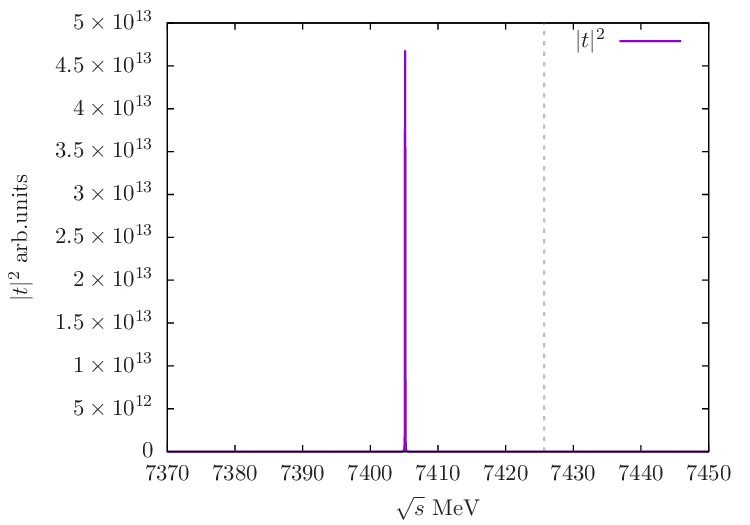}}
\subfigure[$\, |t|^2$ with $q_{\textrm{max}}=500$ ]{
\includegraphics[width=0.3\linewidth]{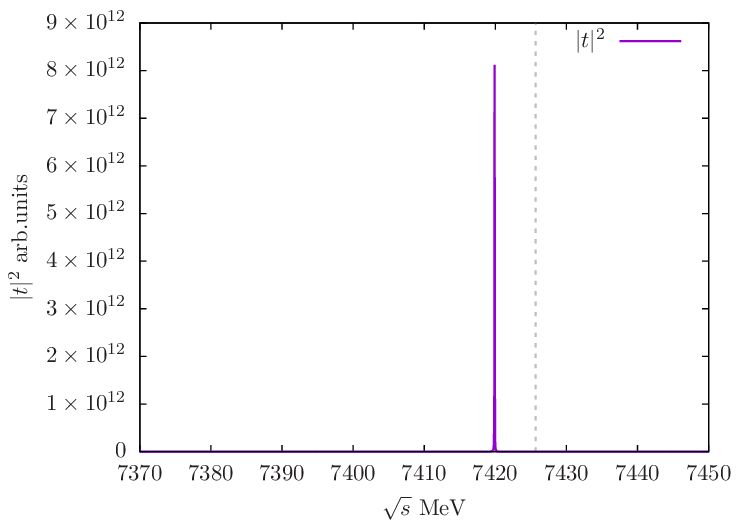}}
\subfigure[$\, |t|^2$ with $q_{\textrm{max}}=400$ ]{
\includegraphics[width=0.3\linewidth]{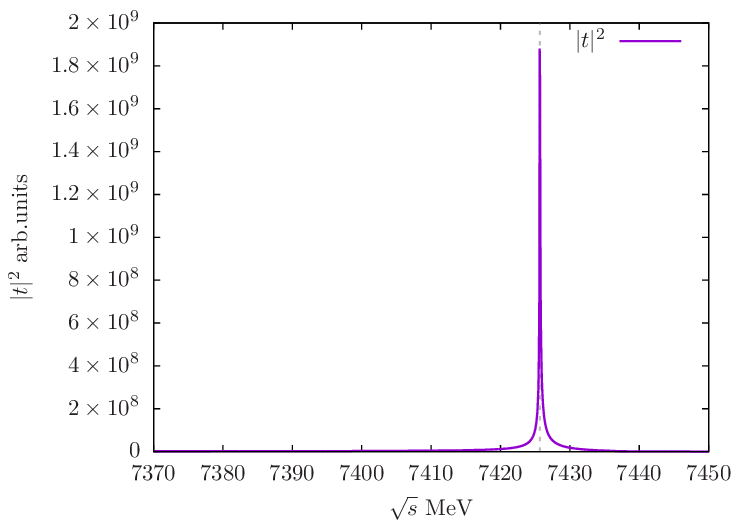}}
\subfigure[$\, |t|^2$ with $q_{\textrm{max}}=300$ ]{
\includegraphics[width=0.3\linewidth]{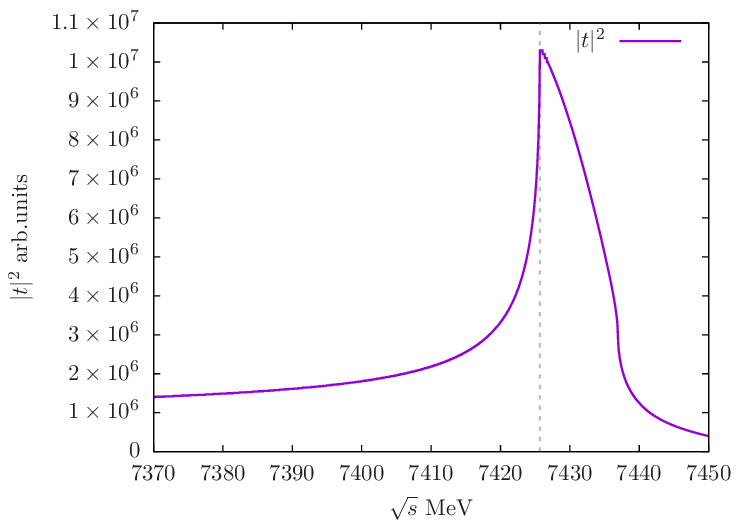}}
\caption{The shape of the transition amplitude $|t(s)|^2 \equiv |T_{VV}^{|(\bar{B}^*D^*)_{s}^-; J=2 \rangle \to |(\bar{B}^*D^*)_{s}^-; J=2 \rangle}(s)|^2$ for the cut-off momentum $q_{\textrm{max}} =$ (a)~700~MeV, (b)~600~MeV, (c)~500~MeV, (d)~400~MeV and (e)~300~MeV. The $\bar B_s^{*0}D^{*+}$ threshold is indicated by a dotted line. The pole is identified as a bound state in the subfigures (a,b,c), while it is identified as a near-threshold virtual state in the subfigures (d,e).}
\label{fig:T with qmax}
\end{figure*}

We use the combination $|(\bar{B}^*D^*)_{s}^-; J=2 \rangle$ as an example, and show its pole position in Fig.~\ref{fig:polepostion1} as a function of the cut-off momentum $q_\textrm{max}$. We find that this pole becomes a bound state when $q_\textrm{max} >410$~MeV, while it becomes a virtual state when $q_\textrm{max} <410$~MeV. We also show its transition amplitude
\begin{equation}
t(s) \equiv T_{VV}^{|(\bar{B}^*D^*)_{s}^-; J=2 \rangle \to |(\bar{B}^*D^*)_{s}^-; J=2 \rangle}(s) \, ,
\end{equation}
in Fig.~\ref{fig:T with qmax} for the cut-off momentum $q_{\textrm{max}} = 700$, $600$, $500$, $400$, and $300$~MeV. This pole is identified as a bound state in Fig.~\ref{fig:T with qmax}(a,b,c), which appears as the singularity under the threshold. Differently, this pole is identified as a virtual state in Fig.~\ref{fig:T with qmax}(d,e), which can significantly enhance the near-threshold cusp effect to produce a sharp peak at the threshold. More discussions on the near-threshold virtual states can be found in Refs.~\cite{Dong:2021bvy,Dong:2020hxe}.

\section{summary}
\label{sec:summary}

In this paper we systematically study the possible hadronic molecular states with the quark contents $bc\bar{s}\bar{q}$, $b\bar{c}s\bar{q}$, and $b\bar{c}\bar{s}q$ ($q=u/d$) within the extended local hidden gauge formalism. We solve the Bethe-Salpeter equation to search for poles on both the physical (first Riemann) sheet and the second Riemann sheet.

We study the $bc\bar{s}\bar{d}$ system by investigating the interactions of the $\bar{B}_s^0D^+$, $\bar{B}^0D_s^+$, $\bar{B}_s^{*0}D^+$, $\bar{B}^{*0}D_s^+$, $\bar{B}_s^0D^{*+}$, $\bar{B}^0D_s^{*+}$, $\bar{B}_s^{*0}D^{*+}$, and $\bar{B}^{*0}D_s^{*+}$ channels. Since the threshold masses of the $\bar{B}_s^0D^+$ and $\bar B^0D_s^+$ channels are quite close to each other, we take into account their mixing as
\begin{eqnarray}
\nonumber |(\bar{B}D)_{s}^+; J=0 \rangle &=& \frac{1}{\sqrt{2}}\left(|\bar{B}_s^0D^+\rangle_{J=0} + |\bar B^0D_s^+\rangle_{J=0}\right),
\\ \nonumber |(\bar{B}D)_{s}^-; J=0 \rangle &=& \frac{1}{\sqrt{2}}\left(|\bar{B}_s^0D^+\rangle_{J=0} - |\bar B^0D_s^+\rangle_{J=0}\right),
\end{eqnarray}
Similar combinations are considered for all the other channels. With the cut-off momentum $q_\textrm{max}=600\mev$, we find six bound states in this system with the binding energies about $10$-$20$~MeV, as summarized in Table~\ref{tab:results1}. These six bound states change to be six near-threshold virtual poles when using the cut-off momentum $q_\textrm{max}=400\mev$, as summarized in Table~\ref{tab:results2}. Similar results are obtained for the $bc\bar{s}\bar{u}$ system, which are also summarized in Tables~\ref{tab:results1} and \ref{tab:results2}.

In the above $bc\bar{s}\bar{q}$ system the interactions are mainly caused by the exchange of the light vector meson $K^*$. However, in the $b\bar{c}s\bar{q}$ and $b\bar{c}\bar{s}q$ systems one can not exchange light vector mesons, and there only the exchanges of the heavy vector mesons $D^*$, $D_s^*$, $B^*$, and $B^*_s$ are allowed. Consequently, the interactions of the $b\bar{c}s\bar{q}$ and $b\bar{c}\bar{s}q$ systems are expected to be significantly smaller, and we do not find any deeply-bound pole in these systems.

To end this paper, we would like to emphasize that the $bc\bar{s}\bar{d}$ and $bc\bar{s}\bar{u}$ systems are rather ``clean'' since there are only limited numbers of coupled channels. We propose to search for these possibly-existing hadronic molecular states in the $\Upsilon$ decays. Besides, we propose to search for $|(\bar{B}^*D^*)_{s}^-; J=2 \rangle$ through its $D$-wave two-body decay patterns $|(\bar{B}^*D^*)_{s}^-; J=2 \rangle \to \bar{B}_s D / \bar B D_s$, $|(\bar{B}^*D^*)_{s}^-; J=1 \rangle$ through its $P$-wave three-body decay patterns $|(\bar{B}^*D^*)_{s}^-; J=1 \rangle \to \bar{B}_sD\pi/ \bar BD_s\pi $, and $|(\bar{B}^*D)_{s}^-; J=1 \rangle$ through its weak decay patterns $|(\bar{B}^*D)_{s}^-; J=1 \rangle \to DD_s\pi$ and its semileptonic decay patterns $|(\bar{B}^*D)_{s}^-; J=1 \rangle \to (D^* D_s/D D_s^*) \ell^-\bar{\nu}_{\ell}$.

\appendix
\section{Interactions in the $b\bar{c}s\bar{q}/b\bar{c}\bar{s}q$ systems}
\label{sec:app}

\begin{figure}[h]
\centering
\includegraphics[width=.9\linewidth]{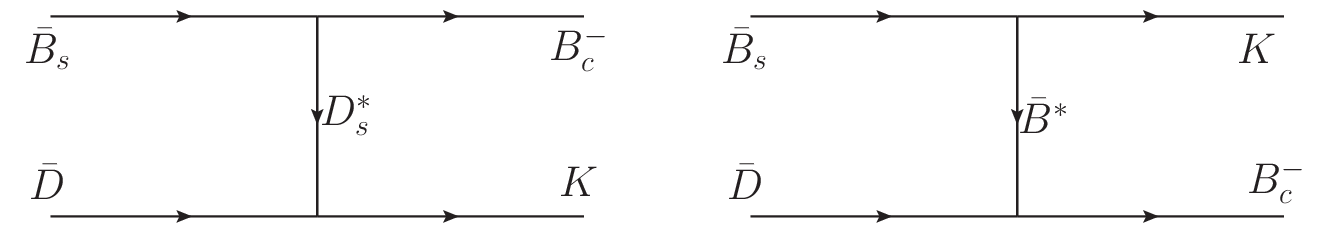}
\caption{The exchange of the vector mesons $D_s^*$ and $\bar B^*$ between two pseudoscalar mesons in the $b\bar{c}s\bar{q}$ system.}
\label{fig-PP-intercation-bcsq}
\end{figure}

\begin{figure}[h]
\centering
\includegraphics[width=.9\linewidth]{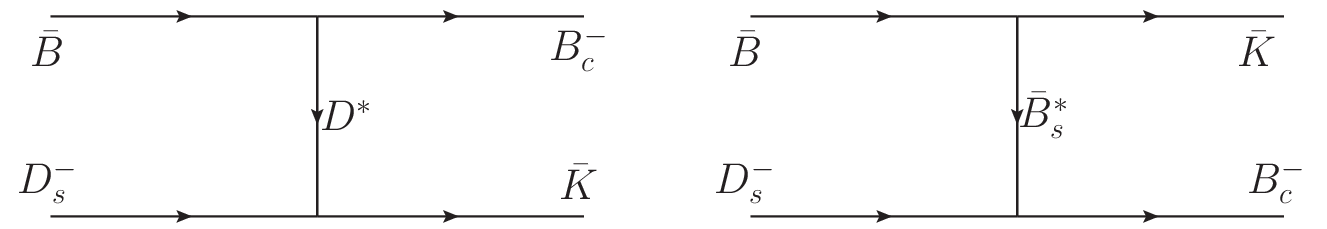}
\caption{The exchange of the vector mesons $D^*$ and $\bar B_s^*$ between two pseudoscalar mesons in the $b\bar{c}\bar{s}q$ system.}
\label{fig-PP-intercation-bcqs}
\end{figure}

In this appendix, we study the interactions of the $b\bar{c}s\bar{q}$ and $b\bar{c}\bar{s}q$ systems. As shown in Fig.~\ref{fig-PP-intercation-bcsq} and Fig.~\ref{fig-PP-intercation-bcqs}, there can only exist the exchange of the heavy vector mesons $D^*$, $D_s^*$, $B^*$, and $B^*_s$, while there does not exist the exchange of the light vector meson $K^*$. Therefore, the interactions of the $b\bar{c}s\bar{q}$ and $b\bar{c}\bar{s}q$ systems are expected to be much smaller than the interaction of the $bc\bar{s}\bar{q}$ system.

When investigating the exchange of the light vector meson $K^*$ in Sec.~\ref{sec:formlism}, we have taken its momentum to be $q^2 \rightarrow 0$ so that its propagator is reduced to be
\begin{eqnarray}
\frac{1}{q^2-m_V^2+i\epsilon} \approx -\frac{1}{m_V^2} \, .
\end{eqnarray}
However, this reduction does not work well when investigating the exchange of heavy mesons, due to the large mass difference between the initial and final mesons. Similar to Ref.~\cite{Yu:2018yxl}, in this paper we adopt the following modification to account for the impact of this effect
\begin{eqnarray}
\frac{1}{(q^0)^2-m_V^2+i\epsilon} \simeq -\lambda\frac{1}{m_V^2},
\end{eqnarray}
where $(q^0)^2=(\Delta M)^2=(M_i-M_j)^2$ and $\Delta M$ is the mass difference between the two external mesons. Since the $B_c^-$ meson is quite massive, we have neglected its three-momentum and approximated the transferred momentum as $q^2 \approx (q^0)^2 = (\Delta M)^2$. Take the processes depicted in Fig.~\ref{fig-PP-intercation-bcsq} as examples, in the left panel $M_i$ and $M_j$ are the masses of the $\bar B_s^0$ and $B_c^-$ mesons, and in the right panel $M_i$ and $M_j$ the masses of the $\bar D$ and $B_c^-$ mesons. Numerically, we obtain
\begin{eqnarray}
\lambda_{\bar B_s \bar D \to B_c^- K}^t &=& \frac{-m_{D_s^*}^2}{(m_{B_c}-m_{B_s})^2-m_{D_s^*}^2}= 1.23, \\
\lambda_{\bar B_s \bar D \to B_c^- K}^u &=& \frac{-m_{B^*}^2}{(m_{B_c}-m_{D})^2-m_{B^*}^2}= 3.17,
\end{eqnarray}
where the superscripts $t$ and $u$ describe the $t$ and $u$ channels, respectively. We note that the contribution from the heavy meson exchange is still quite small compared to the light meson exchange, after the above modifications.

\begin{figure}[h]
\centering
\includegraphics[width=.7\linewidth]{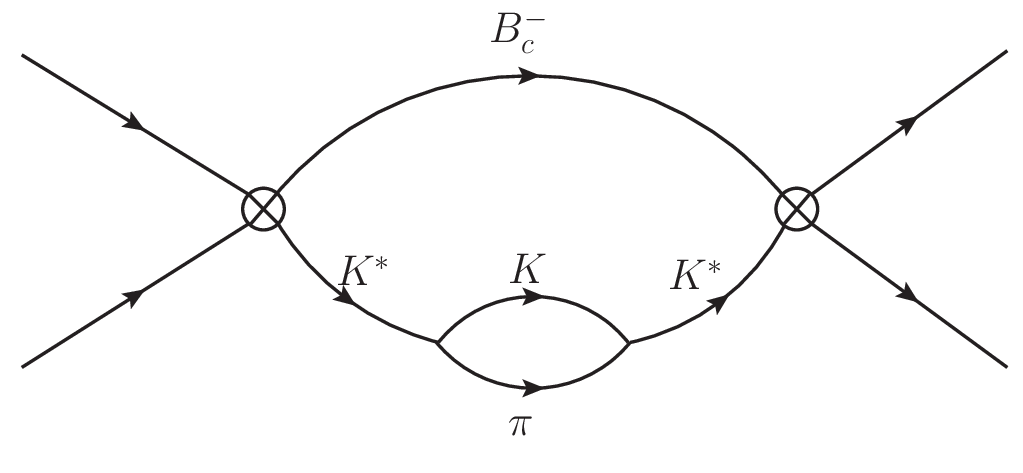}
\caption{The $\bar B_cK^*$ loop taking into account the width of the $K^*$ meson. This provides an imaginary part to the unitarized $\bar B_cK^*$ scattering amplitude.}
\label{fig-KsttoKpi}
\end{figure}

In the calculations we have taken into account the width of the $K^*$ meson, {\it i.e.}, we have taken into account the $K^* \to K \pi$ decay in the $\bar B_cK^*$ loop, as depicted in Fig.~\ref{fig-KsttoKpi}. This provides an imaginary part to the unitarized $\bar B_cK^*$ scattering amplitude, so it can provide a non-zero width to the possibly-existing states generated from the $B_cK^*$ interaction. To take into account this effect in the $B_cK^{*}$ and $B_c^*K^{*}$ channels, we write Eq.~(\ref{eq-cutoff-G}) for the loop function $G(s)$ as
\begin{eqnarray}
\nonumber G(s) &=& \int_0^{q_{\rm max}}\frac{q^2dq}{4\pi^2}\,\frac{\omega_{B_c^{(*)}}+\omega_{K^*}}{\omega_{B_c^{(*)}}\omega_{K^*}} \, \frac{1}{\sqrt{s}+\omega_{B_c^{(*)}}+\omega_{K^*}} \,
\\ && \times \frac{1}{\sqrt{s}-\omega_{K^*}-\omega_{B_c^{(*)}}+i\frac{\sqrt{s'}}{2\omega_{K^*}} \Gamma_{K^*}(s') } \, ,
\label{eq:GwidthK}
\end{eqnarray}
where $s'=(\sqrt{s}-\omega_{B_c^{(*)}})^2-\vec q\,^2$ and
\begin{eqnarray}
\nonumber \Gamma_{K^*}(s') &=& \Gamma_{K^*}(m_{K^*}^2) \frac{m_{K^*}^2}{s'}
\left(\frac{p_\pi(s')}{p_\pi(m_{K^*}^2)}\right)^3
\\ && \times \Theta(\sqrt{s'}-m_K-m_\pi)\,,
\label{eq:widthofsKs}
\end{eqnarray}
with $\Gamma_{K^*}$ the width of the $K^*$ meson.

The rest of the calculations are the same as those used to study the interaction of the $bc\bar{s}\bar{q}$ system. However, for the $b\bar{c}s\bar{q}$ and $b\bar{c}\bar{s}q$ systems, we do not find any pole in both the first and second Riemann sheets, indicating that there do not exist deeply-bound states in these systems due to the weak attraction from the heavy meson exchange.

\section*{Acknowledgments}
%

We are grateful to Eulogio Oset for the very helpful discussion.
This project is supported by
the National Natural Science Foundation of China under Grants No.~12075019 and No.~12192263,
the Jiangsu Provincial Double-Innovation Program under Grant No.~JSSCRC2021488,
the Natural Science Foundation of Henan under Grand No.~222300420554,
the Project of Youth Backbone Teachers of Colleges and Universities of Henan Province (2020GGJS017),
the Youth Talent Support Project of Henan (2021HYTP002),
the Open Project of Guangxi Key Laboratory of Nuclear Physics and Nuclear Technology, No.~NLK2021-08,
and
the Fundamental Research Funds for the Central Universities.

\section*{References}
\bibliographystyle{elsarticle-num}
\bibliography{ref}

\begin{thebibliography}{10}
\expandafter\ifx\csname url\endcsname\relax
  \def\url#1{\texttt{#1}}\fi
\expandafter\ifx\csname urlprefix\endcsname\relax\def\urlprefix{URL }\fi
\expandafter\ifx\csname href\endcsname\relax
  \def\href#1#2{#2} \def\path#1{#1}\fi

\bibitem{LHCb:2021vvq}
R.~Aaij, et~al., {Observation of an exotic narrow doubly charmed tetraquark},
  Nature Phys. 18~(7) (2022) 751--754.
\newblock \href {http://arxiv.org/abs/2109.01038} {\path{arXiv:2109.01038}},
  \href {http://dx.doi.org/10.1038/s41567-022-01614-y}
  {\path{doi:10.1038/s41567-022-01614-y}}.

\bibitem{LHCb:2021auc}
R.~Aaij, et~al., {Study of the doubly charmed tetraquark $T_{cc}^+$}, Nature
  Commun. 13 (2022) 3351.
\newblock \href {http://arxiv.org/abs/2109.01056} {\path{arXiv:2109.01056}},
  \href {http://dx.doi.org/10.1038/s41467-022-30206-w}
  {\path{doi:10.1038/s41467-022-30206-w}}.

\bibitem{Janc:2004qn}
D.~Janc, M.~Rosina, {The $T_{cc} = DD^*$ Molecular State}, Few Body Syst. 35
  (2004) 175--196.
\newblock \href {http://arxiv.org/abs/hep-ph/0405208}
  {\path{arXiv:hep-ph/0405208}}, \href
  {http://dx.doi.org/10.1007/s00601-004-0068-9}
  {\path{doi:10.1007/s00601-004-0068-9}}.

\bibitem{Yang:2009zzp}
Y.~Yang, C.~Deng, J.~Ping, T.~Goldman, {$S$-wave $Q Q \bar q \bar q$ state in
  the constituent quark model}, Phys. Rev. D 80 (2009) 114023.
\newblock \href {http://dx.doi.org/10.1103/PhysRevD.80.114023}
  {\path{doi:10.1103/PhysRevD.80.114023}}.

\bibitem{Li:2012ss}
N.~Li, Z.-F. Sun, X.~Liu, S.-L. Zhu, {Coupled-channel analysis of the possible
  $D^{(*)}D^{(*)}, \overline{B}^{(*)}\overline{B}^{(*)}$ and
  $D^{(*)}\overline{B}^{(*)}$ molecular states}, Phys. Rev. D 88~(11) (2013)
  114008.
\newblock \href {http://arxiv.org/abs/1211.5007} {\path{arXiv:1211.5007}},
  \href {http://dx.doi.org/10.1103/PhysRevD.88.114008}
  {\path{doi:10.1103/PhysRevD.88.114008}}.

\bibitem{Ohkoda:2012hv}
S.~Ohkoda, Y.~Yamaguchi, S.~Yasui, K.~Sudoh, A.~Hosaka, {Exotic mesons with
  double charm and bottom flavor}, Phys. Rev. D 86 (2012) 034019.
\newblock \href {http://arxiv.org/abs/1202.0760} {\path{arXiv:1202.0760}},
  \href {http://dx.doi.org/10.1103/PhysRevD.86.034019}
  {\path{doi:10.1103/PhysRevD.86.034019}}.

\bibitem{Xu:2017tsr}
H.~Xu, B.~Wang, Z.-W. Liu, X.~Liu, {$D D^{*}$ potentials in chiral perturbation
  theory and possible molecular states}, Phys. Rev. D 99~(1) (2019) 014027.
\newblock \href {http://arxiv.org/abs/1708.06918} {\path{arXiv:1708.06918}},
  \href {http://dx.doi.org/10.1103/PhysRevD.99.014027}
  {\path{doi:10.1103/PhysRevD.99.014027}}.

\bibitem{Liu:2019stu}
M.-Z. Liu, T.-W. Wu, M.~Pavon~Valderrama, J.-J. Xie, L.-S. Geng, {Heavy-quark
  spin and flavor symmetry partners of the $X(3872)$ revisited: What can we
  learn from the one boson exchange model?}, Phys. Rev. D 99~(9) (2019) 094018.
\newblock \href {http://arxiv.org/abs/1902.03044} {\path{arXiv:1902.03044}},
  \href {http://dx.doi.org/10.1103/PhysRevD.99.094018}
  {\path{doi:10.1103/PhysRevD.99.094018}}.

\bibitem{Ding:2020dio}
Z.-M. Ding, H.-Y. Jiang, J.~He, {Molecular states from
  $D^{(*)}\bar{D}^{(*)}/B^{(*)}\bar{B}^{(*)}$ and
  $D^{(*)}D^{(*)}/\bar{B}^{(*)}\bar{B}^{(*)}$ interactions}, Eur. Phys. J. C
  80~(12) (2020) 1179.
\newblock \href {http://arxiv.org/abs/2011.04980} {\path{arXiv:2011.04980}},
  \href {http://dx.doi.org/10.1140/epjc/s10052-020-08754-6}
  {\path{doi:10.1140/epjc/s10052-020-08754-6}}.

\bibitem{Qin:2020zlg}
Q.~Qin, Y.-F. Shen, F.-S. Yu, {Discovery potentials of double-charm
  tetraquarks}, Chin. Phys. C 45~(10) (2021) 103106.
\newblock \href {http://arxiv.org/abs/2008.08026} {\path{arXiv:2008.08026}},
  \href {http://dx.doi.org/10.1088/1674-1137/ac1b97}
  {\path{doi:10.1088/1674-1137/ac1b97}}.

\bibitem{BESIII:2020qkh}
M.~Ablikim, et~al., {Observation of a Near-Threshold Structure in the $K^+$
  Recoil-Mass Spectra in $e^+e^- \rightarrow K^+(D_s^-D^{*0}+D_s^{*-}D^0$)},
  Phys. Rev. Lett. 126~(10) (2021) 102001.
\newblock \href {http://arxiv.org/abs/2011.07855} {\path{arXiv:2011.07855}},
  \href {http://dx.doi.org/10.1103/PhysRevLett.126.102001}
  {\path{doi:10.1103/PhysRevLett.126.102001}}.

\bibitem{BESIII:2013ris}
M.~Ablikim, et~al., {Observation of a Charged Charmoniumlike Structure in
  $e^+e^- \to \pi^+ \pi^- J/\psi$ at $\sqrt{s}$ =4.26 GeV}, Phys. Rev. Lett.
  110 (2013) 252001.
\newblock \href {http://arxiv.org/abs/1303.5949} {\path{arXiv:1303.5949}},
  \href {http://dx.doi.org/10.1103/PhysRevLett.110.252001}
  {\path{doi:10.1103/PhysRevLett.110.252001}}.

\bibitem{Belle:2013yex}
Z.~Q. Liu, et~al., {Study of $e^+e^- \to \pi^+ \pi^- J/\psi$ and Observation of
  a Charged Charmoniumlike State at Belle}, Phys. Rev. Lett. 110 (2013) 252002,
  [Erratum: Phys. Rev. Lett. 111, 019901 (2013)].
\newblock \href {http://arxiv.org/abs/1304.0121} {\path{arXiv:1304.0121}},
  \href {http://dx.doi.org/10.1103/PhysRevLett.110.252002}
  {\path{doi:10.1103/PhysRevLett.110.252002}}.

\bibitem{LHCb:2021uow}
R.~Aaij, et~al., {Observation of New Resonances Decaying to $J/\psi K^+$+ and
  $J/\psi \phi$}, Phys. Rev. Lett. 127~(8) (2021) 082001.
\newblock \href {http://arxiv.org/abs/2103.01803} {\path{arXiv:2103.01803}},
  \href {http://dx.doi.org/10.1103/PhysRevLett.127.082001}
  {\path{doi:10.1103/PhysRevLett.127.082001}}.

\bibitem{Lee:2008uy}
S.~H. Lee, M.~Nielsen, U.~Wiedner, {$D_s \bar D^*$ Molecule as an Axial Meson},
  J. Korean Phys. Soc. 55 (2009) 424.
\newblock \href {http://arxiv.org/abs/0803.1168} {\path{arXiv:0803.1168}},
  \href {http://dx.doi.org/10.3938/jkps.55.424}
  {\path{doi:10.3938/jkps.55.424}}.

\bibitem{Ebert:2005nc}
D.~Ebert, R.~N. Faustov, V.~O. Galkin, {Masses of heavy tetraquarks in the
  relativistic quark model}, Phys. Lett. B 634 (2006) 214--219.
\newblock \href {http://arxiv.org/abs/hep-ph/0512230}
  {\path{arXiv:hep-ph/0512230}}, \href
  {http://dx.doi.org/10.1016/j.physletb.2006.01.026}
  {\path{doi:10.1016/j.physletb.2006.01.026}}.

\bibitem{Dias:2013qga}
J.~M. Dias, X.~Liu, M.~Nielsen, {Predicition for the decay width of a charged
  state near the $D_s\bar{D}^*/D^*_s\bar{D}$ threshold}, Phys. Rev. D 88~(9)
  (2013) 096014.
\newblock \href {http://arxiv.org/abs/1307.7100} {\path{arXiv:1307.7100}},
  \href {http://dx.doi.org/10.1103/PhysRevD.88.096014}
  {\path{doi:10.1103/PhysRevD.88.096014}}.

\bibitem{Voloshin:2019ilw}
M.~B. Voloshin, {Strange hadrocharmonium}, Phys. Lett. B 798 (2019) 135022.
\newblock \href {http://arxiv.org/abs/1901.01936} {\path{arXiv:1901.01936}},
  \href {http://dx.doi.org/10.1016/j.physletb.2019.135022}
  {\path{doi:10.1016/j.physletb.2019.135022}}.

\bibitem{Ferretti:2020ewe}
J.~Ferretti, E.~Santopinto, {Hidden-charm and bottom tetra- and pentaquarks
  with strangeness in the hadro-quarkonium and compact tetraquark models}, JHEP
  04 (2020) 119.
\newblock \href {http://arxiv.org/abs/2001.01067} {\path{arXiv:2001.01067}},
  \href {http://dx.doi.org/10.1007/JHEP04(2020)119}
  {\path{doi:10.1007/JHEP04(2020)119}}.

\bibitem{Chen:2013wca}
D.-Y. Chen, X.~Liu, T.~Matsuki, {Predictions of Charged Charmoniumlike
  Structures with Hidden-Charm and Open-Strange Channels}, Phys. Rev. Lett.
  110~(23) (2013) 232001.
\newblock \href {http://arxiv.org/abs/1303.6842} {\path{arXiv:1303.6842}},
  \href {http://dx.doi.org/10.1103/PhysRevLett.110.232001}
  {\path{doi:10.1103/PhysRevLett.110.232001}}.

\bibitem{Chen:2022asf}
H.-X. Chen, W.~Chen, X.~Liu, Y.-R. Liu, S.-L. Zhu, {An updated review of the
  new hadron states}\href {http://arxiv.org/abs/2204.02649}
  {\path{arXiv:2204.02649}}, \href {http://dx.doi.org/10.1088/1361-6633/aca3b6}
  {\path{doi:10.1088/1361-6633/aca3b6}}.

\bibitem{Chen:2016qju}
H.-X. Chen, W.~Chen, X.~Liu, S.-L. Zhu, {The hidden-charm pentaquark and
  tetraquark states}, Phys. Rept. 639 (2016) 1--121.
\newblock \href {http://arxiv.org/abs/1601.02092} {\path{arXiv:1601.02092}},
  \href {http://dx.doi.org/10.1016/j.physrep.2016.05.004}
  {\path{doi:10.1016/j.physrep.2016.05.004}}.

\bibitem{Liu:2019zoy}
Y.-R. Liu, H.-X. Chen, W.~Chen, X.~Liu, S.-L. Zhu, {Pentaquark and Tetraquark
  States}, Prog. Part. Nucl. Phys. 107 (2019) 237--320.
\newblock \href {http://arxiv.org/abs/1903.11976} {\path{arXiv:1903.11976}},
  \href {http://dx.doi.org/10.1016/j.ppnp.2019.04.003}
  {\path{doi:10.1016/j.ppnp.2019.04.003}}.

\bibitem{Lebed:2016hpi}
R.~F. Lebed, R.~E. Mitchell, E.~S. Swanson, {Heavy-quark QCD exotica}, Prog.
  Part. Nucl. Phys. 93 (2017) 143--194.
\newblock \href {http://arxiv.org/abs/1610.04528} {\path{arXiv:1610.04528}},
  \href {http://dx.doi.org/10.1016/j.ppnp.2016.11.003}
  {\path{doi:10.1016/j.ppnp.2016.11.003}}.

\bibitem{Esposito:2016noz}
A.~Esposito, A.~Pilloni, A.~D. Polosa, {Multiquark resonances}, Phys. Rept. 668
  (2017) 1--97.
\newblock \href {http://arxiv.org/abs/1611.07920} {\path{arXiv:1611.07920}},
  \href {http://dx.doi.org/10.1016/j.physrep.2016.11.002}
  {\path{doi:10.1016/j.physrep.2016.11.002}}.

\bibitem{Hosaka:2016pey}
A.~Hosaka, T.~Iijima, K.~Miyabayashi, Y.~Sakai, S.~Yasui, {Exotic hadrons with
  heavy flavors: X, Y, Z, and related states}, PTEP 2016~(6) (2016) 062C01.
\newblock \href {http://arxiv.org/abs/1603.09229} {\path{arXiv:1603.09229}},
  \href {http://dx.doi.org/10.1093/ptep/ptw045}
  {\path{doi:10.1093/ptep/ptw045}}.

\bibitem{Guo:2017jvc}
F.-K. Guo, C.~Hanhart, U.-G. Mei\ss{}ner, Q.~Wang, Q.~Zhao, B.-S. Zou,
  {Hadronic molecules}, Rev. Mod. Phys. 90~(1) (2018) 015004.
\newblock \href {http://arxiv.org/abs/1705.00141} {\path{arXiv:1705.00141}},
  \href {http://dx.doi.org/10.1103/RevModPhys.90.015004}
  {\path{doi:10.1103/RevModPhys.90.015004}}.

\bibitem{Ali:2017jda}
A.~Ali, J.~S. Lange, S.~Stone, {Exotics: Heavy pentaquarks and tetraquarks},
  Prog. Part. Nucl. Phys. 97 (2017) 123--198.
\newblock \href {http://arxiv.org/abs/1706.00610} {\path{arXiv:1706.00610}},
  \href {http://dx.doi.org/10.1016/j.ppnp.2017.08.003}
  {\path{doi:10.1016/j.ppnp.2017.08.003}}.

\bibitem{Olsen:2017bmm}
S.~L. Olsen, T.~Skwarnicki, D.~Zieminska, {Nonstandard heavy mesons and
  baryons: Experimental evidence}, Rev. Mod. Phys. 90~(1) (2018) 015003.
\newblock \href {http://arxiv.org/abs/1708.04012} {\path{arXiv:1708.04012}},
  \href {http://dx.doi.org/10.1103/RevModPhys.90.015003}
  {\path{doi:10.1103/RevModPhys.90.015003}}.

\bibitem{Karliner:2017qhf}
M.~Karliner, J.~L. Rosner, T.~Skwarnicki, {Multiquark States}, Ann. Rev. Nucl.
  Part. Sci. 68 (2018) 17--44.
\newblock \href {http://arxiv.org/abs/1711.10626} {\path{arXiv:1711.10626}},
  \href {http://dx.doi.org/10.1146/annurev-nucl-101917-020902}
  {\path{doi:10.1146/annurev-nucl-101917-020902}}.

\bibitem{Bass:2018xmz}
S.~D. Bass, P.~Moskal, {$\eta^\prime$ and $\eta$ mesons with connection to
  anomalous glue}, Rev. Mod. Phys. 91~(1) (2019) 015003.
\newblock \href {http://arxiv.org/abs/1810.12290} {\path{arXiv:1810.12290}},
  \href {http://dx.doi.org/10.1103/RevModPhys.91.015003}
  {\path{doi:10.1103/RevModPhys.91.015003}}.

\bibitem{Brambilla:2019esw}
N.~Brambilla, S.~Eidelman, C.~Hanhart, A.~Nefediev, C.-P. Shen, C.~E. Thomas,
  A.~Vairo, C.-Z. Yuan, {The $XYZ$ states: Experimental and theoretical status
  and perspectives}, Phys. Rept. 873 (2020) 1--154.
\newblock \href {http://arxiv.org/abs/1907.07583} {\path{arXiv:1907.07583}},
  \href {http://dx.doi.org/10.1016/j.physrep.2020.05.001}
  {\path{doi:10.1016/j.physrep.2020.05.001}}.

\bibitem{Guo:2019twa}
F.-K. Guo, X.-H. Liu, S.~Sakai, {Threshold cusps and triangle singularities in
  hadronic reactions}, Prog. Part. Nucl. Phys. 112 (2020) 103757.
\newblock \href {http://arxiv.org/abs/1912.07030} {\path{arXiv:1912.07030}},
  \href {http://dx.doi.org/10.1016/j.ppnp.2020.103757}
  {\path{doi:10.1016/j.ppnp.2020.103757}}.

\bibitem{Ketzer:2019wmd}
B.~Ketzer, B.~Grube, D.~Ryabchikov, {Light-meson spectroscopy with COMPASS},
  Prog. Part. Nucl. Phys. 113 (2020) 103755.
\newblock \href {http://arxiv.org/abs/1909.06366} {\path{arXiv:1909.06366}},
  \href {http://dx.doi.org/10.1016/j.ppnp.2020.103755}
  {\path{doi:10.1016/j.ppnp.2020.103755}}.

\bibitem{Yang:2020atz}
G.~Yang, J.~Ping, J.~Segovia, {Tetra- and Penta-Quark Structures in the
  Constituent Quark Model}, Symmetry 12~(11) (2020) 1869.
\newblock \href {http://arxiv.org/abs/2009.00238} {\path{arXiv:2009.00238}},
  \href {http://dx.doi.org/10.3390/sym12111869}
  {\path{doi:10.3390/sym12111869}}.

\bibitem{Roberts:2021nhw}
C.~D. Roberts, D.~G. Richards, T.~Horn, L.~Chang, {Insights into the emergence
  of mass from studies of pion and kaon structure}, Prog. Part. Nucl. Phys. 120
  (2021) 103883.
\newblock \href {http://arxiv.org/abs/2102.01765} {\path{arXiv:2102.01765}},
  \href {http://dx.doi.org/10.1016/j.ppnp.2021.103883}
  {\path{doi:10.1016/j.ppnp.2021.103883}}.

\bibitem{Fang:2021wes}
S.-S. Fang, B.~Kubis, A.~Kup{\'s}{\'c}, {What can we learn about light-meson
  interactions at electron-positron colliders?}, Prog. Part. Nucl. Phys. 120
  (2021) 103884.
\newblock \href {http://arxiv.org/abs/2102.05922} {\path{arXiv:2102.05922}},
  \href {http://dx.doi.org/10.1016/j.ppnp.2021.103884}
  {\path{doi:10.1016/j.ppnp.2021.103884}}.

\bibitem{Jin:2021vct}
S.~Jin, X.~Shen, {Highlights of light meson spectroscopy at the BESIII
  experiment}, Natl. Sci. Rev. 8~(11) (2021) nwab198.
\newblock \href {http://dx.doi.org/10.1093/nsr/nwab198}
  {\path{doi:10.1093/nsr/nwab198}}.

\bibitem{JPAC:2021rxu}
M.~Albaladejo, et~al., {Novel approaches in hadron spectroscopy}, Prog. Part.
  Nucl. Phys. 127 (2022) 103981.
\newblock \href {http://arxiv.org/abs/2112.13436} {\path{arXiv:2112.13436}},
  \href {http://dx.doi.org/10.1016/j.ppnp.2022.103981}
  {\path{doi:10.1016/j.ppnp.2022.103981}}.

\bibitem{Meng:2022ozq}
L.~Meng, B.~Wang, G.-J. Wang, S.-L. Zhu, {Chiral perturbation theory for heavy
  hadrons and chiral effective field theory for heavy hadronic molecules}\href
  {http://arxiv.org/abs/2204.08716} {\path{arXiv:2204.08716}}.

\bibitem{Mai:2022eur}
M.~Mai, U.-G. Mei\ss{}ner, C.~Urbach, {Towards a theory of hadron
  resonances}\href {http://arxiv.org/abs/2206.01477} {\path{arXiv:2206.01477}}.

\bibitem{Maiani:2022psl}
L.~Maiani, A.~Pilloni, {GGI Lectures on Exotic Hadrons}, 2022.
\newblock \href {http://arxiv.org/abs/2207.05141} {\path{arXiv:2207.05141}}.

\bibitem{Ling:2021bir}
X.-Z. Ling, M.-Z. Liu, L.-S. Geng, E.~Wang, J.-J. Xie, {Can we understand the
  decay width of the $T_{cc}^+$ state?}, Phys. Lett. B 826 (2022) 136897.
\newblock \href {http://arxiv.org/abs/2108.00947} {\path{arXiv:2108.00947}},
  \href {http://dx.doi.org/10.1016/j.physletb.2022.136897}
  {\path{doi:10.1016/j.physletb.2022.136897}}.

\bibitem{Chen:2020aos}
H.-X. Chen, W.~Chen, R.-R. Dong, N.~Su, {$X_0$(2900) and $X_1$(2900): Hadronic
  Molecules or Compact Tetraquarks}, Chin. Phys. Lett. 37~(10) (2020) 101201.
\newblock \href {http://arxiv.org/abs/2008.07516} {\path{arXiv:2008.07516}},
  \href {http://dx.doi.org/10.1088/0256-307X/37/10/101201}
  {\path{doi:10.1088/0256-307X/37/10/101201}}.

\bibitem{Chen:2020uif}
H.-X. Chen, W.~Chen, X.~Liu, X.-H. Liu, {Establishing the first hidden-charm
  pentaquark with strangeness}, Eur. Phys. J. C 81~(5) (2021) 409.
\newblock \href {http://arxiv.org/abs/2011.01079} {\path{arXiv:2011.01079}},
  \href {http://dx.doi.org/10.1140/epjc/s10052-021-09196-4}
  {\path{doi:10.1140/epjc/s10052-021-09196-4}}.

\bibitem{Chen:2021erj}
H.-X. Chen, {Hadronic molecules in $B$ decays}, Phys. Rev. D 105~(9) (2022)
  094003.
\newblock \href {http://arxiv.org/abs/2103.08586} {\path{arXiv:2103.08586}},
  \href {http://dx.doi.org/10.1103/PhysRevD.105.094003}
  {\path{doi:10.1103/PhysRevD.105.094003}}.

\bibitem{Meissner:1987ge}
U.~G. Meissner, {Low-Energy Hadron Physics from Effective Chiral Lagrangians
  with Vector Mesons}, Phys. Rept. 161 (1988) 213.
\newblock \href {http://dx.doi.org/10.1016/0370-1573(88)90090-7}
  {\path{doi:10.1016/0370-1573(88)90090-7}}.

\bibitem{Bando:1987br}
M.~Bando, T.~Kugo, K.~Yamawaki, {Nonlinear Realization and Hidden Local
  Symmetries}, Phys. Rept. 164 (1988) 217--314.
\newblock \href {http://dx.doi.org/10.1016/0370-1573(88)90019-1}
  {\path{doi:10.1016/0370-1573(88)90019-1}}.

\bibitem{Nagahiro:2008cv}
H.~Nagahiro, L.~Roca, A.~Hosaka, E.~Oset, {Hidden gauge formalism for the
  radiative decays of axial-vector mesons}, Phys. Rev. D 79 (2009) 014015.
\newblock \href {http://arxiv.org/abs/0809.0943} {\path{arXiv:0809.0943}},
  \href {http://dx.doi.org/10.1103/PhysRevD.79.014015}
  {\path{doi:10.1103/PhysRevD.79.014015}}.

\bibitem{Wu:2010jy}
J.-J. Wu, R.~Molina, E.~Oset, B.~S. Zou, {Prediction of Narrow $N^*$ and
  $\Lambda^*$ Resonances with Hidden Charm above 4~GeV}, Phys. Rev. Lett. 105
  (2010) 232001.
\newblock \href {http://arxiv.org/abs/1007.0573} {\path{arXiv:1007.0573}},
  \href {http://dx.doi.org/10.1103/PhysRevLett.105.232001}
  {\path{doi:10.1103/PhysRevLett.105.232001}}.

\bibitem{Duan:2021pll}
M.-Y. Duan, G.-Y. Wang, E.~Wang, D.-M. Li, D.-Y. Chen, {Revisiting the $Z_c$
  (4025) structure observed by BESIII in $e^+e^- \to (D^* \bar
  D^*)^{\pm,0}\pi^{\mp,0}$ at $\sqrt {s}$ =4.26 GeV}, Phys. Rev. D 104~(7)
  (2021) 074030.
\newblock \href {http://arxiv.org/abs/2109.00731} {\path{arXiv:2109.00731}},
  \href {http://dx.doi.org/10.1103/PhysRevD.104.074030}
  {\path{doi:10.1103/PhysRevD.104.074030}}.

\bibitem{Zhang:2020rqr}
Y.~Zhang, E.~Wang, D.-M. Li, Y.-X. Li, {Search for the $D^*\bar{D}^*$ molecular
  state $Z_c(4000)$ in the reaction $B^{-} \rightarrow J/\psi \rho^0 K^{-}$},
  Chin. Phys. C 44~(9) (2020) 093107.
\newblock \href {http://arxiv.org/abs/2001.06624} {\path{arXiv:2001.06624}},
  \href {http://dx.doi.org/10.1088/1674-1137/44/9/093107}
  {\path{doi:10.1088/1674-1137/44/9/093107}}.

\bibitem{Wang:2017mrt}
E.~Wang, J.-J. Xie, L.-S. Geng, E.~Oset, {Analysis of the $B^+\to J/\psi \phi
  K^+$ data at low $J/\psi \phi$ invariant masses and the $X(4140)$ and
  $X(4160)$ resonances}, Phys. Rev. D 97~(1) (2018) 014017.
\newblock \href {http://arxiv.org/abs/1710.02061} {\path{arXiv:1710.02061}},
  \href {http://dx.doi.org/10.1103/PhysRevD.97.014017}
  {\path{doi:10.1103/PhysRevD.97.014017}}.

\bibitem{Geng:2008gx}
L.~S. Geng, E.~Oset, {Vector meson-vector meson interaction in a hidden gauge
  unitary approach}, Phys. Rev. D 79 (2009) 074009.
\newblock \href {http://arxiv.org/abs/0812.1199} {\path{arXiv:0812.1199}},
  \href {http://dx.doi.org/10.1103/PhysRevD.79.074009}
  {\path{doi:10.1103/PhysRevD.79.074009}}.

\bibitem{Xiao:2013yca}
C.~W. Xiao, J.~Nieves, E.~Oset, {Combining heavy quark spin and local hidden
  gauge symmetries in the dynamical generation of hidden charm baryons}, Phys.
  Rev. D 88 (2013) 056012.
\newblock \href {http://arxiv.org/abs/1304.5368} {\path{arXiv:1304.5368}},
  \href {http://dx.doi.org/10.1103/PhysRevD.88.056012}
  {\path{doi:10.1103/PhysRevD.88.056012}}.

\bibitem{Molina:2008jw}
R.~Molina, D.~Nicmorus, E.~Oset, {The rho rho interaction in the hidden gauge
  formalism and the f(0)(1370) and f(2)(1270) resonances}, Phys. Rev. D 78
  (2008) 114018.
\newblock \href {http://arxiv.org/abs/0809.2233} {\path{arXiv:0809.2233}},
  \href {http://dx.doi.org/10.1103/PhysRevD.78.114018}
  {\path{doi:10.1103/PhysRevD.78.114018}}.

\bibitem{Uchino:2015uha}
T.~Uchino, W.-H. Liang, E.~Oset, {Baryon states with hidden charm in the
  extended local hidden gauge approach}, Eur. Phys. J. A 52~(3) (2016) 43.
\newblock \href {http://arxiv.org/abs/1504.05726} {\path{arXiv:1504.05726}},
  \href {http://dx.doi.org/10.1140/epja/i2016-16043-0}
  {\path{doi:10.1140/epja/i2016-16043-0}}.

\bibitem{Oller:1997ti}
J.~A. Oller, E.~Oset, {Chiral symmetry amplitudes in the S wave isoscalar and
  isovector channels and the $\sigma$, f$_0$(980), a$_0$(980) scalar mesons},
  Nucl. Phys. A 620 (1997) 438--456, [Erratum: Nucl.Phys.A 652, 407--409
  (1999)].
\newblock \href {http://arxiv.org/abs/hep-ph/9702314}
  {\path{arXiv:hep-ph/9702314}}, \href
  {http://dx.doi.org/10.1016/S0375-9474(97)00160-7}
  {\path{doi:10.1016/S0375-9474(97)00160-7}}.

\bibitem{Oller:1998hw}
J.~A. Oller, E.~Oset, J.~R. Pelaez, {Meson meson interaction in a
  nonperturbative chiral approach}, Phys. Rev. D 59 (1999) 074001, [Erratum:
  Phys.Rev.D 60, 099906 (1999), Erratum: Phys.Rev.D 75, 099903 (2007)].
\newblock \href {http://arxiv.org/abs/hep-ph/9804209}
  {\path{arXiv:hep-ph/9804209}}, \href
  {http://dx.doi.org/10.1103/PhysRevD.59.074001}
  {\path{doi:10.1103/PhysRevD.59.074001}}.

\bibitem{Oller:1997ng}
J.~A. Oller, E.~Oset, J.~R. Pelaez, {Nonperturbative approach to effective
  chiral Lagrangians and meson interactions}, Phys. Rev. Lett. 80 (1998)
  3452--3455.
\newblock \href {http://arxiv.org/abs/hep-ph/9803242}
  {\path{arXiv:hep-ph/9803242}}, \href
  {http://dx.doi.org/10.1103/PhysRevLett.80.3452}
  {\path{doi:10.1103/PhysRevLett.80.3452}}.

\bibitem{Oset:1997it}
E.~Oset, A.~Ramos, {Nonperturbative chiral approach to s wave anti-K N
  interactions}, Nucl. Phys. A 635 (1998) 99--120.
\newblock \href {http://arxiv.org/abs/nucl-th/9711022}
  {\path{arXiv:nucl-th/9711022}}, \href
  {http://dx.doi.org/10.1016/S0375-9474(98)00170-5}
  {\path{doi:10.1016/S0375-9474(98)00170-5}}.

\bibitem{Jido:2003cb}
D.~Jido, J.~A. Oller, E.~Oset, A.~Ramos, U.~G. Meissner, {Chiral dynamics of
  the two Lambda(1405) states}, Nucl. Phys. A 725 (2003) 181--200.
\newblock \href {http://arxiv.org/abs/nucl-th/0303062}
  {\path{arXiv:nucl-th/0303062}}, \href
  {http://dx.doi.org/10.1016/S0375-9474(03)01598-7}
  {\path{doi:10.1016/S0375-9474(03)01598-7}}.

\bibitem{Sakai:2017avl}
S.~Sakai, L.~Roca, E.~Oset, {Charm-beauty meson bound states from
  $B(B^*)D(D^*)$ and $B(B^*)\bar D(\bar D^*)$ interaction}, Phys. Rev. D 96~(5)
  (2017) 054023.
\newblock \href {http://arxiv.org/abs/1704.02196} {\path{arXiv:1704.02196}},
  \href {http://dx.doi.org/10.1103/PhysRevD.96.054023}
  {\path{doi:10.1103/PhysRevD.96.054023}}.

\bibitem{Dai:2022ulk}
L.~R. Dai, E.~Oset, A.~Feijoo, R.~Molina, L.~Roca, A.~M. Torres, K.~P.
  Khemchandani, {Masses and widths of the exotic molecular $B_{(s)}^{(*)}
  B_{(s)}^{(*)}$ states}, Phys. Rev. D 105~(7) (2022) 074017.
\newblock \href {http://arxiv.org/abs/2201.04840} {\path{arXiv:2201.04840}},
  \href {http://dx.doi.org/10.1103/PhysRevD.105.074017}
  {\path{doi:10.1103/PhysRevD.105.074017}}.

\bibitem{Oset:2022xji}
E.~Oset, L.~Roca, {Exotic molecular meson states of $B^{(*)} K^{(*)}$ nature},
  Eur. Phys. J. C 82~(10) (2022) 882.
\newblock \href {http://arxiv.org/abs/2207.08538} {\path{arXiv:2207.08538}},
  \href {http://dx.doi.org/10.1140/epjc/s10052-022-10850-8}
  {\path{doi:10.1140/epjc/s10052-022-10850-8}}.

\bibitem{Debastiani:2017ewu}
V.~R. Debastiani, J.~M. Dias, W.~H. Liang, E.~Oset, {Molecular $\Omega_c$
  states generated from coupled meson-baryon channels}, Phys. Rev. D 97~(9)
  (2018) 094035.
\newblock \href {http://arxiv.org/abs/1710.04231} {\path{arXiv:1710.04231}},
  \href {http://dx.doi.org/10.1103/PhysRevD.97.094035}
  {\path{doi:10.1103/PhysRevD.97.094035}}.

\bibitem{Feijoo:2021ppq}
A.~Feijoo, W.~H. Liang, E.~Oset, {$D^0D^0\pi^+$ mass distribution in the
  production of the $T_{cc}$ exotic state}, Phys. Rev. D 104~(11) (2021)
  114015.
\newblock \href {http://arxiv.org/abs/2108.02730} {\path{arXiv:2108.02730}},
  \href {http://dx.doi.org/10.1103/PhysRevD.104.114015}
  {\path{doi:10.1103/PhysRevD.104.114015}}.

\bibitem{Dong:2021bvy}
X.-K. Dong, F.-K. Guo, B.-S. Zou, {A survey of heavy-heavy hadronic molecules},
  Commun. Theor. Phys. 73~(12) (2021) 125201.
\newblock \href {http://arxiv.org/abs/2108.02673} {\path{arXiv:2108.02673}},
  \href {http://dx.doi.org/10.1088/1572-9494/ac27a2}
  {\path{doi:10.1088/1572-9494/ac27a2}}.

\bibitem{Dong:2020hxe}
X.-K. Dong, F.-K. Guo, B.-S. Zou, {Explaining the Many Threshold Structures in
  the Heavy-Quark Hadron Spectrum}, Phys. Rev. Lett. 126~(15) (2021) 152001.
\newblock \href {http://arxiv.org/abs/2011.14517} {\path{arXiv:2011.14517}},
  \href {http://dx.doi.org/10.1103/PhysRevLett.126.152001}
  {\path{doi:10.1103/PhysRevLett.126.152001}}.

\bibitem{Yu:2018yxl}
Q.~X. Yu, R.~Pavao, V.~R. Debastiani, E.~Oset, {Description of the $\Xi_c$ and
  $\Xi_b$ states as molecular states}, Eur. Phys. J. C 79~(2) (2019) 167.
\newblock \href {http://arxiv.org/abs/1811.11738} {\path{arXiv:1811.11738}},
  \href {http://dx.doi.org/10.1140/epjc/s10052-019-6665-z}
  {\path{doi:10.1140/epjc/s10052-019-6665-z}}.

\end{thebibliography}

\end{document}